\documentclass[aps,onecolumn,superscriptaddress]{revtex4-1}
\usepackage{lineno}

\usepackage{graphicx,amsmath,verbatim}

\newcommand{\transp}{\dag}
\newcommand{\dd}{\mathrm{d}}
\renewcommand{\figurename}{Fig.}

\usepackage{color}

\newcommand{\red}[1]{#1}

\begin{document}

\title{A novel framework to analyze complex network dynamics}

\author{M. Gilson}
\affiliation{Center for Brain and Cognition, Department of Information and Communication Technologies, Universitat Pompeu Fabra, Carrer Ramon Trias Fargas, 25-27, 08005 Barcelona, Spain.}

\author{N. E. Kouvaris}
\affiliation{Center for Brain and Cognition, Department of Information and Communication Technologies, Universitat Pompeu Fabra, Carrer Ramon Trias Fargas, 25-27, 08005 Barcelona, Spain.}
\affiliation{Namur Institute for Complex Systems (naXys), Department of Mathematics, University of Namur, Rempart de la Vierge 8, B 5000 Namur, Belgium}

\author{G. Deco}
\affiliation{Center for Brain and Cognition, Department of Information and Communication Technologies, Universitat Pompeu Fabra, Carrer Ramon Trias Fargas, 25-27, 08005 Barcelona, Spain.}
\affiliation{Instituci{\'o} Catalana de la Recerca i Estudis Avanats (ICREA), Universitat Pompeu Fabra, Passeig Llu{\'i}s Companys 23, Barcelona, 08010, Spain}

\author{G. Zamora-L{\'o}pez}
\affiliation{Center for Brain and Cognition, Department of Information and Communication Technologies, Universitat Pompeu Fabra, Carrer Ramon Trias Fargas, 25-27, 08005 Barcelona, Spain.}

\begin{abstract}

Graph theory constitutes a widely used and established field providing powerful tools for the characterization of complex networks. The intricate topology of networks can also be investigated by means of the collective dynamics observed in the interactions of self-sustained oscillations (synchronization patterns) or propagation-like processes such as random walks. However, networks are often inferred from real data forming dynamic systems, which are different from those employed to reveal their topological characteristics. This stresses the necessity for a theoretical framework dedicated to the mutual relationship between the structure and dynamics in complex networks, as the two sides of the same coin. Here we propose a rigorous framework based on the network response over time (i.e., Green function) to study interactions between nodes across time. For this purpose we define the \emph{flow} that describes the interplay between the network connectivity and external inputs. This multivariate measure relates to the concepts of graph communicability and the map equation. We illustrate our theory using the multivariate Ornstein-Uhlenbeck process, which describes stable and non-conservative dynamics, but the formalism can be adapted to other local dynamics for which the Green function is known. We provide applications to classical network examples, such as small-world ring and hierarchical networks.
Our theory defines a comprehensive framework that is canonically related to directed and weighted networks, thus paving a new way to revise the standards for network analysis, from the pairwise interactions between nodes to the global properties of networks including community detection.

\end{abstract}

\maketitle

\section{Introduction}

The study of complex networks has become a central tool to investigate many natural and man-made systems in various scientific and technical domains, such as sociology~\cite{Borgatti2009}, neuroscience~\cite{Sporns2000, Sporns2013}, biology~\cite{Jeong2000}, chemistry~\cite{Wickramasinghe2013, Kouvaris2016} and telecommunications~\cite{Broder2000}. As a descendant of classical graph theory, the primary toolbox to study complex networks relies on statistical descriptors like the distribution of degrees, clustering coefficient and centrality of nodes~\cite{Albert2002, Newman2010}. Initially designed for symmetric binary graphs, these measures have been extended to investigate directed~\cite{Bang2009} and weighted~\cite{Barrat2004} networks, aiming to interpret real-world data. While accounting for the directed nature of links is rather straight-forward, the study of weighted networks with ``off-the-shelf'' metrics inherited from graph theory is less natural. In real networks the weights associated to the links represent physical or statistical quantities, beyond the mere existence or absence of the link. Therefore, predefined measures and formulae for binary graphs are often limited, which underlines the need for formalisms that are better suited for the study and interpretation of weighted networks.

The mutual relationship between network structure and dynamics has been studied in both directions. On the one hand, intricate topologies support the emergence of complex collective dynamics in networks~\cite{Boccaletti2006, Arenas2008}. 
The description of networks using the graph measures provides intuitive, but \emph{largely simplified} information about how the network topology may affect its dynamics. 
For example, strongly connected clusters of nodes are expected to synchronize internally before synchronizing with each other.
As an effort to link the network structure to the pairwise functional associations of nodes,~\citet{Estrada2008} introduced \emph{communicability}. The rationale behind is to take into account indirect paths in addition to direct paths in the network in order to evaluate the interactions between nodes.
This measure was used to assess the contribution of structural topology to functional connectivity in fMRI data~\cite{Bettinardi2017}.

On the other hand, the behavior of multivariate network dynamics has been employed to reveal the structural organization of complex topologies~\cite{Arenas2006, Boccaletti2007, Rosvall2008}. 
Connectivity patterns, from the local to global scales, induce a variety of timescales in the functional interactions between nodes and groups thereof. 
Accordingly, the multivariate Ornstein-Uhlenbeck (MOU) process was used to define the notion of network complexity, relying on the entropy of the correlation pattern resulting from a given network connectivity~\cite{Tononi1994, Galan2008, Barnett_PRE_2009}.
Another direction, based on the collective dynamics of coupled phase oscillators, was developed to reveal communities and hierarchical scales along the path to global synchrony since denser structures synchronize before sparser components~\cite{Arenas2006}. 
A related approach exploited the diffusion of random walkers in graphs to reveal community structure: The map equation searches the simplest description of the random walks in a two-stage hierarchy that defines communities~\cite{Rosvall2008}.

However, the bidirectional relationship between topology and dynamics is rarely studied simultaneously, i.e., considering ``the two sides of the same coin''. Moreover, one aspect of the data analysis is often overlooked: Many real networks are inferred from multivariate signals that have a temporal structure. This means that these data should be interpreted as a dynamic system, taking time into account.
Here we strive to reach an overarching viewpoint and define graph-like measures for such complex network dynamics.

In the present study, we develop a theoretical framework to characterize and explore the properties of complex network dynamics. It is based on the MOU process~\cite{Lutkepohl2005}, which can be interpreted as a non-conservative propagation of fluctuating activity in a network with linear feedback~\cite{Gilson_BiolCyb_2017}. 
The MOU process has been used for a long time to study the Brownian motion~\cite{Doob1942} and applied to model data in many fields, such as interest growth in economy~\cite{Vasicek1977}, epidemic spreading~\cite{Britton2010, Andersson2011} and fMRI in neuroscience~\cite{Gilson2016, Gilson2017}.
It has also been used to quantify network complexity~\cite{Galan2008, Barnett_PRE_2009}.
From the theory of linearly coupled dynamics we derive two core measures: \emph{dynamic communicability} and \emph{flow}, which serve as the basis for multivariate network descriptors. 
These are tightly related to the Green function of the coupled MOU process ---the matrix exponential of the Jacobian for the MOU dynamics--- that describes the network response resulting from a unitary impulse at a given node.
This framework is canonically related to directed and weighted networks, aiming to lift limitations of tools derived from current graph theory.
\red{It sits in the general context of matrix exponentials applied on adjacency matrix or their Laplacian to explore graph properties~\cite{Estrada2007, Estrada2012, Schaub2012, Mugnolo2017}.}

The manuscript is organized as follows. In Section~\ref{sec2} we introduce the novel framework and illustrate it with simple network examples. There, we contrast our theory to previously proposed formalisms of communicability~\cite{Estrada2008, Estrada2013}, community detection~\cite{Arenas2006, Schaub2012} \red{and heat kernel~\cite{Chung1992, Berard1994}}. We introduce a novel network metric that quantifies the heterogeneity of interactions resulting from the dynamic communicability or flow, which we term \emph{diversity}.
Section~\ref{sec3} presents three applications of our framework to stereotypical synthetic networks. The first two examples deal with the properties of random graphs and small-world networks~\cite{Watts1998}, while the third example illustrates the potential of our formalism to detect community structure and hierarchical levels in networks~\cite{Rosvall2008, Schaub2012}. The manuscript finishes with a last example from dynamic systems (not graph theory), which examines balanced dynamics in a network with excitatory and inhibitory nodes.

\section{Theory for stable network dynamics with linear feedback} \label{sec2}

In this section we introduce a framework to characterize the properties of complex networks through induced dynamics on their topology. To do so, we consider the multivariate Ornstein-Uhlenbeck process ---a non-conservative and stable propagation of fluctuating activity--- and develop graph-like measures to describe the relationship between connectivity and network dynamics. 
First, we define \emph{dynamic communicability} to characterize the impulse response of the network due to its connectivity.
Meanwhile, we relate our theory to a previously proposed \red{formalisms that also involve matrix exponentials to quantify interactions or relationships between nodes in networks}.
Then, we take into account the effect of inputs with the definition of the \emph{flow}, of which dynamic communicability is a particular case.
The section ends with an investigation of the spectral properties of the flow and the definition of the measure of \emph{diversity}, which is used with the applications to classical benchmark networks in Section~\ref{sec3}.

\begin{figure*}
\includegraphics[scale=1]{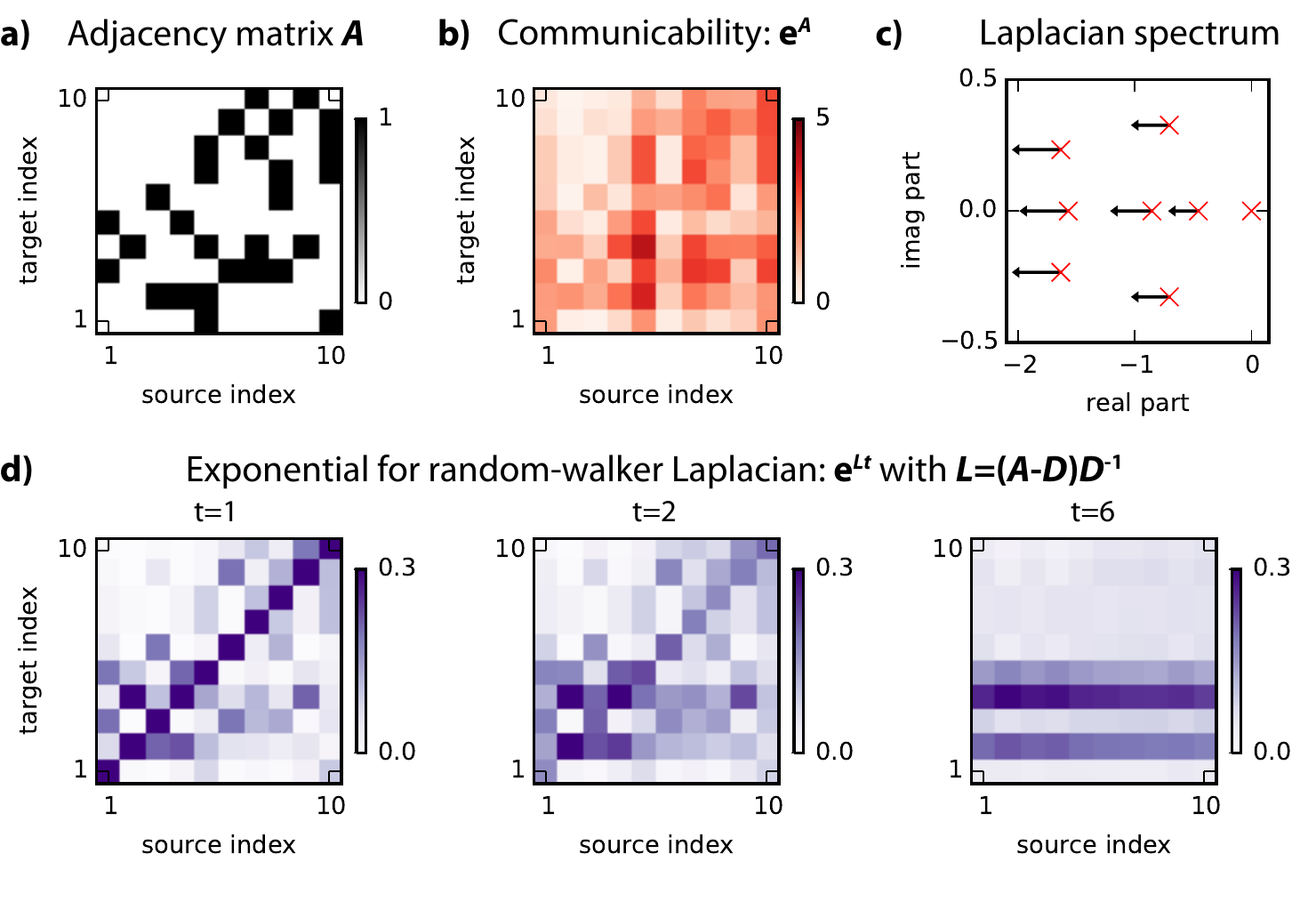}
\caption{ \label{fig1_intro}
{\bf Communicability and Laplacian flow for ``static'' graphs.}
{\bf a)} Binary directed adjacency matrix $A$.
{\bf b)} Communicability defined as $e^A$~\cite{Estrada2008}.
{\bf c)} Schematic evolution of the eigenvalues of the random-walk Laplacian $L = (A-D) D^{-1}$ in Eq.~\eqref{eq:Laplacian} with $D$ being a diagonal matrix with the degree of each node~\cite{Schaub2012}. 
Arrows indicate the ``speed'' of the corresponding eigenvalues, which are larger for more negative real parts.
{\bf d)} Exponential matrix of the random-walk Laplacian $e^{L t}$ at three abstract ``time'' snapshots.
}
\end{figure*}

The concept of communicability for graphs was proposed by~\citet{Estrada2008} to evaluate the influence that nodes exert over one another relying on two simple, but realistic assumptions. 
This measure postulates that (i) the interaction between nodes accumulates along all possible paths of various lengths, not only the shortest paths; and that (ii) shorter paths are more influential than longer paths.
In practice, given the adjacency matrix $A$ of a network, communicability is defined as the matrix exponential of the adjacency matrix, $e^A$, see \figurename~\ref{fig1_intro}a and b. Since the matrix exponential has an exact series expansion $e^A = \sum_{n \geq 0} A^n / n!$, communicability can be understood as a summation of influence over all possible paths with a factorial decay for the influence of the paths (given by the powers $A^n$) depending on their lengths $n$. 
Although communicability has been related to the Green function \red{or Hamiltonian} of a network of coupled springs~\cite{Estrada2012, Estrada2013}, its precise dynamical interpretation has remained rather unclear. 
In Annex~\ref{an:comm}, we show a rigorous formalization based on a cascade of activity in a network~\cite{Motter2002}, for which $e^A$ corresponds to the growth rate for activity in continuous time. Such a system is non-conservative as each node sends a ``unit of activity'' to all its targets for each unit received, so the total activity on the network rapidly grows over time and diverges. 
This definition can be extended to examine weighted and directed adjacency matrices~\cite{Estrada2013}, as shown in the example $A$ in \figurename~\ref{fig1_intro}.

The above definition of communicability is suitable to study graphs, but limited for complex networks associated with many real dynamic systems.
To show this point, we consider the MOU process that has been used to model such network dynamics in many scientific disciplines~\cite{Doob1942, Vasicek1977, Britton2010, Andersson2011, Gilson2016, Gilson2017}.
A MOU process is determined by (i) a local leakage, (ii) a directed weighted graph associated with linear coupling and (iii) input covariances.
It describes the propagation of activity over a network:
	\begin{equation} 	\label{eq:MOU}
	\dd x_i = \left( - \frac{x_i}{\tau_i} + \sum_{1 \leq j \leq N} A_{ij} \, x_j \right) \dd t + \dd z_i \, ,
	\end{equation}
where $\tau_i$ is a decay time constant for node $i$, $A_{ij}$ is the connection weight from node $j$ to node $i$, and $z_i$ is a Wiener process representing the fluctuating input received by node $i$. 
If we ignore the dissipation due to the local leakage $\tau_i$ and the noisy inputs $z_i$, the system reduces to the non-conservative exploding cascading system, see Annex~\ref{an:comm} with Eq.~\eqref{eq:cascade} related to the above-mentioned graph communicability.
Intuitively, stability requires $\tau_i$ to be sufficiently large such that the dissipation at the nodes is faster than the growth due to the cascading effect determined by the connectivity.
In matrix form, Eq.~\eqref{eq:MOU} can be written as 
	\begin{equation} 	\label{eq:MOU_mat}
	\dd \mathbf{x} = J \mathbf{x} \dd t \, + \, \dd \mathbf{z} \, ,
	\end{equation}
where the Jacobian matrix $J$ is determined by the leakage time constant and the connectivity as
	\begin{equation} 	\label{eq:Jacobian}
	J_{ij} = - \frac{\delta_{ij}}{\tau_i} + A_{ij} \, ,
	\end{equation}
where $\delta_{ij}$ is the Kronecker delta.
Notice that we employ the usual convention in dynamical systems rather than that of graph theory: $A_{ij}$ is the weight of the link from node $j$ to node $i$.
In the following we will consider in most cases that nodes have identical $\tau_i = \tau$. 

The solution of Eq.~\eqref{eq:MOU} has a canonical relationship with the matrix exponential of its Jacobian~\cite{Lutkepohl2005}. 
Given the initial conditions $\mathbf{x}(0)$ at time $t = 0$, the state at time $t>0$ is given by~\citet{Lutkepohl2005}
	\begin{equation} 	\label{eq:int_act}
	\mathbf{x}(t) = e^{J t} \; \mathbf{x}(0) + \int_0^t\! e^{J u} \; \dd \mathbf{z}^{t-u} \, ,
	\end{equation}
which also depends on the particular realization of $\mathbf{z}$.
Interestingly, the contribution of the connectivity on the activity of the nodes for a time interval $\Delta$ is quantified by the matrix exponential $e^{J \Delta}$ for both contributions: The element $(i,j)$ of $e^{J t}$ describes the effect of the impulse response from $j$ onto $i$ after time $t$ when taking network effects into account ---corresponding to the Green function of the ordinary linear differential system in Eq.~\eqref{eq:MOU}. The communicability proposed by~\citet{Estrada2008} thus corresponds to $t = 1$ and ignores the temporal evolution of the matrix when $t$ varies, as well as the dissipation due to the diagonal matrix elements.

The matrix exponential is also reminiscent of the formalism developed to examine the hierarchical structure of Kuramoto oscillators~\cite{Arenas2006} and of the map equation~\cite{Schaub2012} for complex graphs, where the graph Laplacian $L$ replaces the Jacobian $J$. 
In those studies, a spectral analysis of $e^{L t}$ while varying the (abstract) time $t$ reveals a hierarchical community structure.
This phenomenon is illustrated for the map equation in \figurename~\ref{fig1_intro}c: The zero eigenvalue remains still, while all other eigenvalues are eliminated towards the left side (starting with those that have the largest negative real parts). 
Therefore, fewer and fewer eigenvalues determine the network structure of $e^{L t}$, which becomes simpler and eventually converges towards a row matrix.
The transition from $t = 1$ to $t = 6$ in \figurename~\ref{fig1_intro}c can be used to determine communities: Increasing $t$ spans the hierarchies in the graph and allows for a multiscale description of the graph in \figurename~\ref{fig1_intro}a. The overall structure has simplified from $t = 1$ to $t = 2$, indicating a possible community structure corresponding to nodes with similar rows. 
For the row matrix at $t = 6$, all columns of $e^{L t}$ become very close to the stationary distribution of random walkers, which depends solely on $L$. Details about the mathematical formulation are provided in Annex~\ref{an:map_eq}, see Eq.~\eqref{eq:Laplacian} for the dynamic system giving rise to $e^{L t}$ in Eq.~\eqref{eq:exp_Lap} and the spectral decomposition in Eq.~\eqref{eq:decomp_Laplacian}.

There are \red{four} important differences (some being related to one another) that are worth stressing between our approach and previous work:
\begin{enumerate}
    \item
    The dynamic regime is non-conservative and stable for the MOU, which is suitable to study many real dynamic systems where time has a natural and concrete meaning.
    The local leakage determined by $\tau_i$ is equivalent to a negative self-connection for each node, such that the Jacobian $J$ has eigenvalues with strictly negative real part for the Jacobian $J$.
    For a positive coupling matrix $A$ and identical $\tau_i = \tau$, the dominating eigenvalue (or spectral diameter) of $A$ needs to satisfy $\lambda_\mathrm{max} < -1/\tau$ in order to counterbalance the global feedback determined by $A$ (faster decay than the cascading growth process).
    In comparison, the dynamic system related to the graph communicability in Eq.~\eqref{eq:cascade} is non-conservative, but not stable (exploding).
    On another hand, the map equation in Eq.~\eqref{eq:Laplacian} and \red{heat kernels correspond to conservative systems, as they involve each a type of Laplacian that has a zero eigenvalue~\cite{Chung1992, Berard1994, Schaub2012}}.
    \item
    The basis of our framework is the network response over time, which corresponds to the Green function and is the basis of the concept of dynamic communicability.
    \red{Because of the stable nature of the MOU dynamics, the network response decays to zero (moreover, it is integrable). This means that the interest is on the temporal evolution of the node interactions described by the Green function.
    In contrast, previous studies give a ``static'' picture of node interactions~\cite{Estrada2008}, even for temporal networks~\cite{Estrada2013}. 
    This comes from a description based on a dynamic system related to an ``imaginary time'' or abstract ``inverse temperature'', which diverges when the inverse temperature increases (so the analysis focuses on given finite times).}
    \item
    \red{Another difference concerns the ``normalization'' associated with the matrix exponential.
    The use of the Laplacian~\cite{Chung1992, Berard1994, Schaub2012, Estrada2012} can be seen as a normalized version by the node degrees, compared to graph communicability~\cite{Estrada2008, Estrada2013} and exponential of the adjacency matrix in general~\cite{Mugnolo2017}.
    In our case we rely on the subtraction of the matrix exponential for the corresponding unconnected network (similar to a null model). The rationale is to evaluate the extra contribution due to the connectivity, as will be specified below.}
    \item
    The MOU dynamics are also determined by the input properties, in addition to the network connectivity. 
    Unlike the equilibrium distribution $\tilde{\mathbf{x}}$ of random walkers ---see Eq.~\eqref{eq:stationary_distrib} in Annex\ref{an:map_eq}--- the inputs are \emph{independent} from the connectivity. 
    Moreover, the variable of interest for those inputs $z_i$ is their covariance matrix $\Sigma$, that is, their second-order statistics.
    This follows because the dynamic system defined by Eq.~\eqref{eq:MOU} is dissipative, so its activity fades to the same fixed point irrespective of the initial condition $\mathbf{x}(0)$. 
    We thus focus on the operation of the network connectivity on the input covariance matrix $\Sigma$, which shapes the covariances of the node activities $x_i$~\cite{Gilson_BiolCyb_2017}. 
    This will be the basis of the concept of flow.
\end{enumerate}
\red{Together, this points to a richer description of the stable dynamic MOU system, incorporating the temporal dimension and input properties.}

\subsection{Dynamic communicability as a measure of interactions across time}

\begin{figure*}
\includegraphics[scale=1]{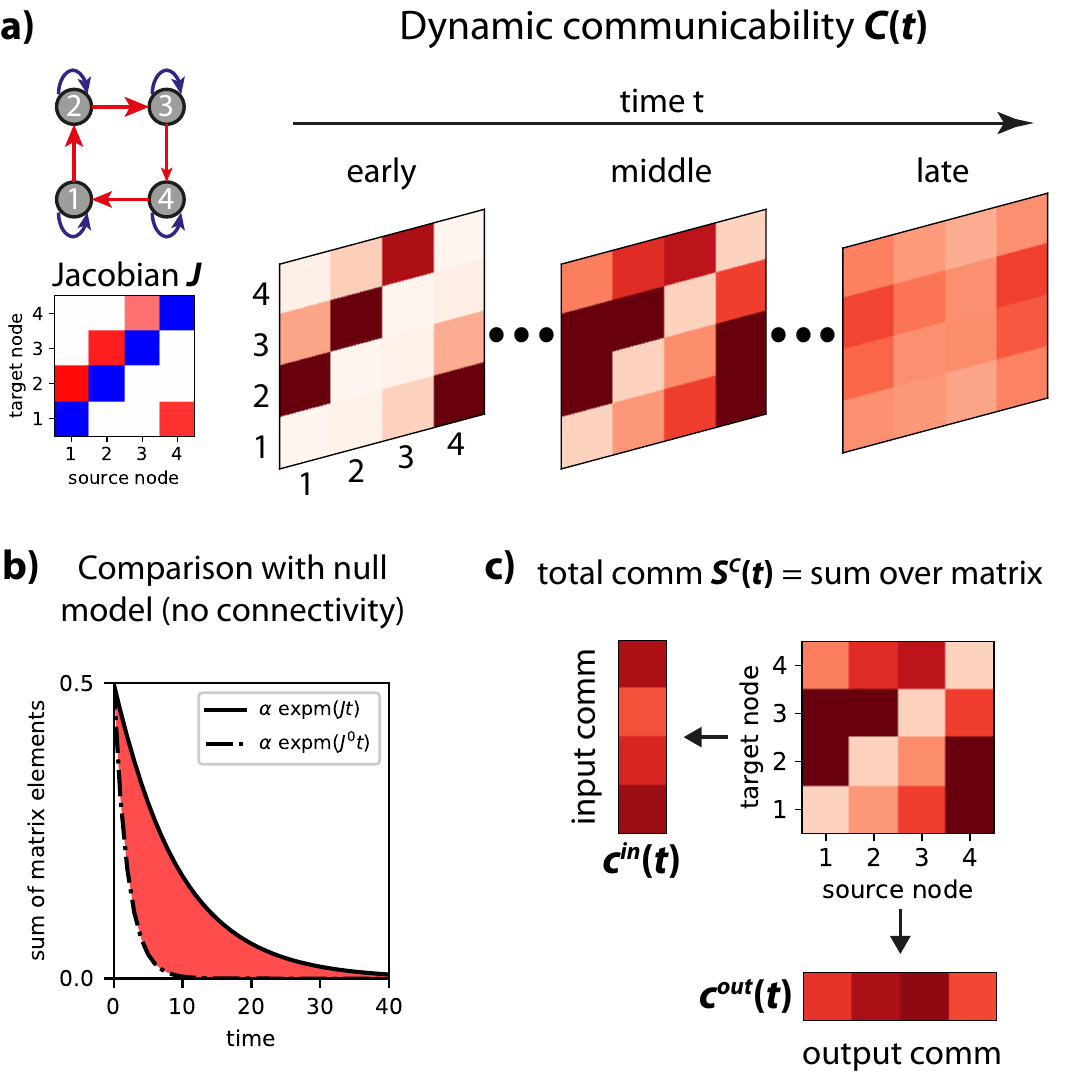}
\caption{ \label{fig2_def_comm}
{\bf Dynamic communicability for a multivariate Ornstein-Uhlenbeck (MOU) process.}
{\bf a)} The dynamics for the network (top left) is determined by the Jacobian matrix $J$ (bottom left) with $-1/\tau_i$ on the matrix diagonal in blue and the recurrent connectivity $A$ (off-diagonal elements) in red. 
Dynamic communicability is the family of matrices $\mathcal{C}(t)$ in Eq.~\eqref{eq:def_comm}, which involves the exponential of the Jacobian multiplied by time $t$.
{\bf b)} The sum $\mathcal{S}^\mathcal{C}(t)$ of all matrix elements of $\mathcal{C}(t)$ quantifies the total effect due to the recurrent connections (red area between the curves). Here the scaling factor is $\alpha = || J^0 ||$.
{\bf c)} The sums of the matrix elements in $\mathcal{C}(t)$ along rows and columns give the input and output communicability for each node, respectively. 
}
\end{figure*}

Now we focus on the activity propagation through to the recurrent connectivity in the MOU process to define several time-dependent metrics that characterize the influence of the topology on the network dynamics, ignoring the input properties. \emph{Dynamic communicability} is the ``deformation'' of the Green function of the MOU, namely $e^{J t}$ in Eq.~\eqref{eq:int_act}, due to the presence of the connectivity embodied by the (weighted and directed) matrix $A$. As mentioned above, this is quantified by the following subtraction with a MOU process that has the same leakage $J^0_{ij} = -\delta_{ij}/\tau_i$, but no connectivity:
        \begin{equation} \label{eq:def_comm}
        \mathcal{C}(t) = || J^0 || \big( e^{J t} - e^{J^0 t} \big) \, .
        \end{equation}
The scaling factor $|| J^0 ||^{-1} = || \int_{t \geq 0} e^{J^0 t} \dd t ||$ is used for normalization purpose; $|| \cdot ||$ is the L1-norm for matrices (i.e., sum of elements in absolute value). We coin the measure with the term ``dynamic'' to stress that the matrix $\mathcal{C}(t)$ evolves over time as illustrated by the successive matrices $\mathcal{C}(t)$ in \figurename~\ref{fig2_def_comm}a. 

From the matrix family $\mathcal{C}(t)$ ---akin to a space-space-time tensor--- we define several simplified measure to interpret the information, while keeping the focus on the temporal evolution.
The total communicability $\mathcal{S}^\mathcal{C}(t)$ is the sum of all elements of $\mathcal{C}(t)$ at a given time $t$:
        \begin{equation} \label{eq:def_tot_comm}
	\mathcal{S}^\mathcal{C}(t) = \mathbf{e}^\transp \mathcal{C}(t) \mathbf{e} = \sum_{1 \leq i,j \leq N} \mathcal{C}(t) \, ,
        \end{equation}
We have used the unit vector $\mathbf{e}$ of dimension $N$. 
The presence of connections $A_{ij} > 0$ increases the values of the matrix $e^{J t}$, whose sum is represented by the black curve in \figurename~\ref{fig2_def_comm}b, to be compared with the dashed-dotted curve for $e^{J^0 t}$.
The difference between the curves (red area) gives $\mathcal{S}^\mathcal{C}(t)$. 

Following the tradition of graph theory, which provides metrics to characterize the properties of a network at different scales, we can evaluate the properties of individual nodes. \red{As done previously with graph communicability~\cite{Estrada2013}}, we define the input and the output communicability of a node as the row and column sums respectively: 
        \begin{eqnarray} \label{eq:def_inout_comm}
        \mathbf{c}^\mathrm{in}(t) & = & \mathcal{C}(t) \; \mathbf{e} \, ,
        \\
        \mathbf{c}^\mathrm{out}(t) & = & \mathcal{C}(t)^\transp \; \mathbf{e} \, .
        \nonumber
        \end{eqnarray}
The input and output communicabilities of the example network in \figurename~\ref{fig2_def_comm}a are shown in \figurename~\ref{fig2_def_comm}c. Notice that the vector elements of $\mathbf{c}^\mathrm{in}(t)$ and $\mathbf{c}^\mathrm{out}(t)$  ---with vector elements $\sum_{1 \leq j \leq N} \mathcal{C}_{ij}(t)$ and $\sum_{1 \leq j \leq N} \mathcal{C}_{ji}(t)$ indexed by $1 \leq i \leq N$, respectively--- are also time-dependent measures and thus they allow us to investigate how the properties of a node evolves from the moment the network has been perturbed until activity stops due to the dominant dispersion term.

Finally, the integration of the matrices $\mathcal{C}(t)$ over time gives the overall interaction between two nodes over the whole time $t \geq 0$, yielding the matrix
        \begin{equation} \label{eq:time_integr_comm}
        \overline{\mathcal{C}} = \int_0^\infty \mathcal{C}(t) \, \dd t = || J^0 || \big[ J^{-1} - (J^0)^{-1} \big] \, ,
        \end{equation}
where the inverse of the Jacobian appears.
In the following, we distinguish semantically between the concrete connections in the network (embodied in $A$) and the interactions resulting from the propagation of activity in the network, which are quantified by the dynamic communicability. Note that nodes without direct connection have a non-zero interaction if there exists at least a path allowing them to communicate via other nodes.

\begin{figure*}
\includegraphics[scale=1]{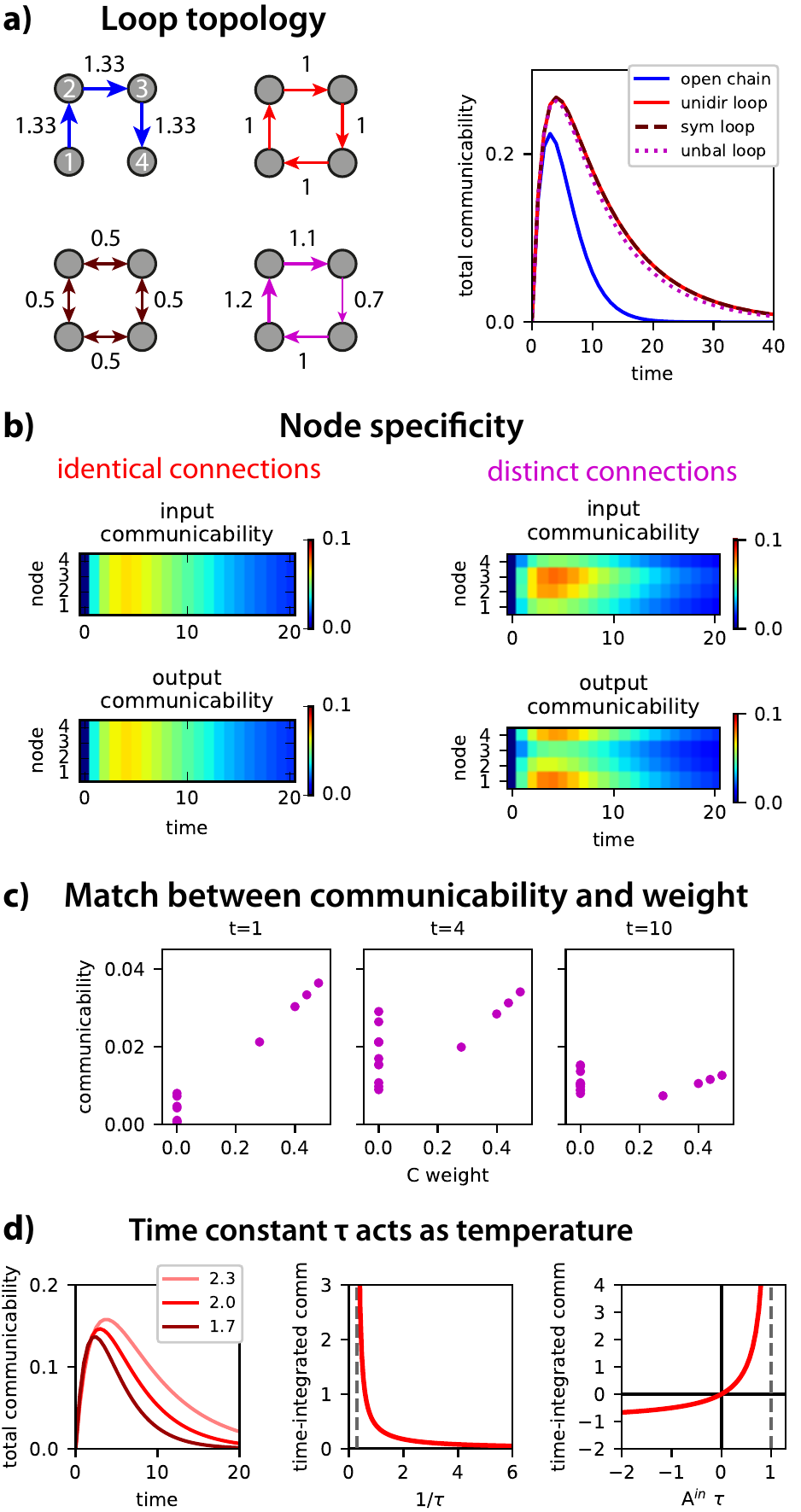}
\caption{ \label{fig3_loop}
{\bf Spatiotemporal properties of dynamic communicability.}
{\bf a)} Example network topologies (left) with colored links (self loops are not represented) and their total communicability over time (right).
{\bf b)} Input and output communicabilities for the loop networks with identical and distinct weights (red and magenta links in panel {\bf a}).
{\bf c)} Correspondence between the weight and communicability for each pair of nodes at three time snapshots for the network with unbalanced weights (magenta links in panel {\bf a}).
{\bf d)} The left panel shows the total communicability over time for three different values of $\tau$ as shown in the legend. 
The middle panel displays the theoretical relationship between the total communicability integrated over time $\overline{\mathcal{S}^\mathcal{C}}$ in Eq.~\eqref{eq:tint_comm_av} as a function of the $1/\tau$ for a given $A^\mathrm{in}$. 
The time constant $\tau$ acts as a ``temperature'' that reduces $\overline{\mathcal{S}^\mathcal{C}}$ for small values (i.e., high values of $1/\tau$); note also that $\overline{\mathcal{S}^\mathcal{C}}$ diverges when $\tau$ reaches the value $1/A^\mathrm{in}$ (dashed line). 
The right panel shows a similar plot with the compound parameter $A^\mathrm{in} \tau$ appearing in Eq.~\eqref{eq:tint_comm_av} on the x-axis.
}
\end{figure*}

To illustrate how dynamic communicability captures the properties of the network topology, we examine four simple networks, see \figurename~\ref{fig3_loop}a. Blue arrows correspond to a directed open chain ($1 \rightarrow 2 \rightarrow 3 \rightarrow 4$). Red and magenta arrows to a directed closed loop ($1 \rightarrow 2 \rightarrow 3 \rightarrow 4 \rightarrow 1$) with equal and distinct weights, respectively. Last, brown arrows to a bidirectional cycle.
The corresponding matrices $A$ have the same total weight.
Right panel of \figurename~\ref{fig3_loop}a, shows that the total communicability for the loop networks is larger than for the open chain, even though the weights sum equally in the connectivity matrices. 
This can be understood by expanding the matrix exponential as a series of matrices corresponding to direct connections, then paths of lengths 2, 3 and so on
        \begin{equation} \label{eq:decomp_exp}
        e^{J t} = e^{-t/\tau} e^{A t} = e^{-t/\tau} \sum_{n \geq 0} \frac{t^n A^n}{n!} \ ,
        \end{equation}
where we have used the fact that the $\tau_i = \tau$ are all equal.
For the open chain we have that $A^n = 0$ for $n \geq N = 4$, whereas these matrices have positive elements for loops and contribute to the total communicability for all interactions, including those corresponding to nodes that are not directly connected.
This results in larger sustained total communicability for the recurrent topologies than for the open chain.

Although the total communicability is almost identical for the three loop networks, the nodes have differentiated roles, as captured by the input and output communicabilities defined in Eq.~\eqref{eq:def_inout_comm} and displayed in \figurename~\ref{fig3_loop}b: The  responses are the same for all nodes in the red loop with equal weights, but exhibit variety for the magenta loop with distinct weights. 
The temporal evolution of the communicability matrices thus convey important information about the roles of nodes. Due to the asymmetry in input and output connections some nodes seem to play the role of ``broadcasting'' information while others are clearly more likely to act as ``receivers''. \figurename~\ref{fig3_loop}c displays the evolution of the communicability for individual connections: It is initially aligned with the connection strengths (left panel), then becomes more homogeneous especially with an increase of communicability for unconnected nodes (middle panel) before fading out eventually (right panel).

Finally, we briefly look into the role of the time constant $\tau$ on the diagonal of $J$, as well as that of $J^0$.
The mean-field approximation lumps nodes together as a single unit with self-feedback $-1 / \tau + A^\mathrm{in}$, where $\tau_i = \tau$ is assumed to be identical and $A^\mathrm{in}$ is the mean input weight to each node.
In this way, the total communicability and its time-integrated value can be approximated by
        \begin{eqnarray}
        \mathcal{S}^\mathcal{C}(t) & \simeq & \frac{e^{-t/\tau} (e^{A^\mathrm{in} t} - 1)}{\tau} \ ,
        \label{eq:tot_comm_av}
        \\
        \overline{\mathcal{S}^\mathcal{C}} & \simeq & \int_0^{+\infty} \frac{e^{-t/\tau} (e^{A^\mathrm{in} t} - 1)}{\tau} \dd t = \frac{A^\mathrm{in} \tau}{1 - A^\mathrm{in} \tau} \ .
        \label{eq:tint_comm_av}
        \end{eqnarray}
Here the factor $|| J^0 ||$ in Eq.~\eqref{eq:def_comm} leads to a homogeneous dimensionless formulation, as well as normalizes communicability regardless of the network size.
The time constant $\tau$ thus acts as a temperature, as can be seen in \figurename~\ref{fig3_loop}d: Large values for $\tau$ correspond to increased $\mathcal{S}^\mathcal{C}(t)$ in the left panels and $\overline{\mathcal{S}^\mathcal{C}}$ in the middle panel. 
Compared to Estrada's communicability $e^A$, we do not choose whether the relevant time is, e.g. $t = 1$ or $t = \tau$, which give different matrix structures $e^{-1/\tau} e^{A}$ and $e^{-1} e^{A \tau}$. 
Instead, we consider the whole time line as done in the Laplacian formalization of the map equation~\cite{Schaub2012}.
The temperature-like effect of $\tau$ is also reminiscent of the extension of the concept of communicability to a multivariate autoregressive process of order larger than 2~\cite{Estrada2013}: An ``inverse temperature'' parameter is introduced to determine a scaling factor between discrete time steps.
The MOU dynamics is stable for $A^\mathrm{in} < \tau^{-1}$, which corresponds to exploding communicability for too strong network feedback, as indicated in the middle and right panels when $A^\mathrm{in}$ reaches $\tau^{-1}$ or $A^\mathrm{in} \tau$ reaches 1 (vertical dashed gray lines).
In other words, a unit quantity of activity homogeneously injected in the network corresponds to $\overline{\mathcal{S}^\mathcal{C}}$ on average after circulating through the nodes.

\subsection{Definition of flow to quantify the propagation of fluctuating inputs via the network connectivity}

\begin{figure*}
\includegraphics[scale=1]{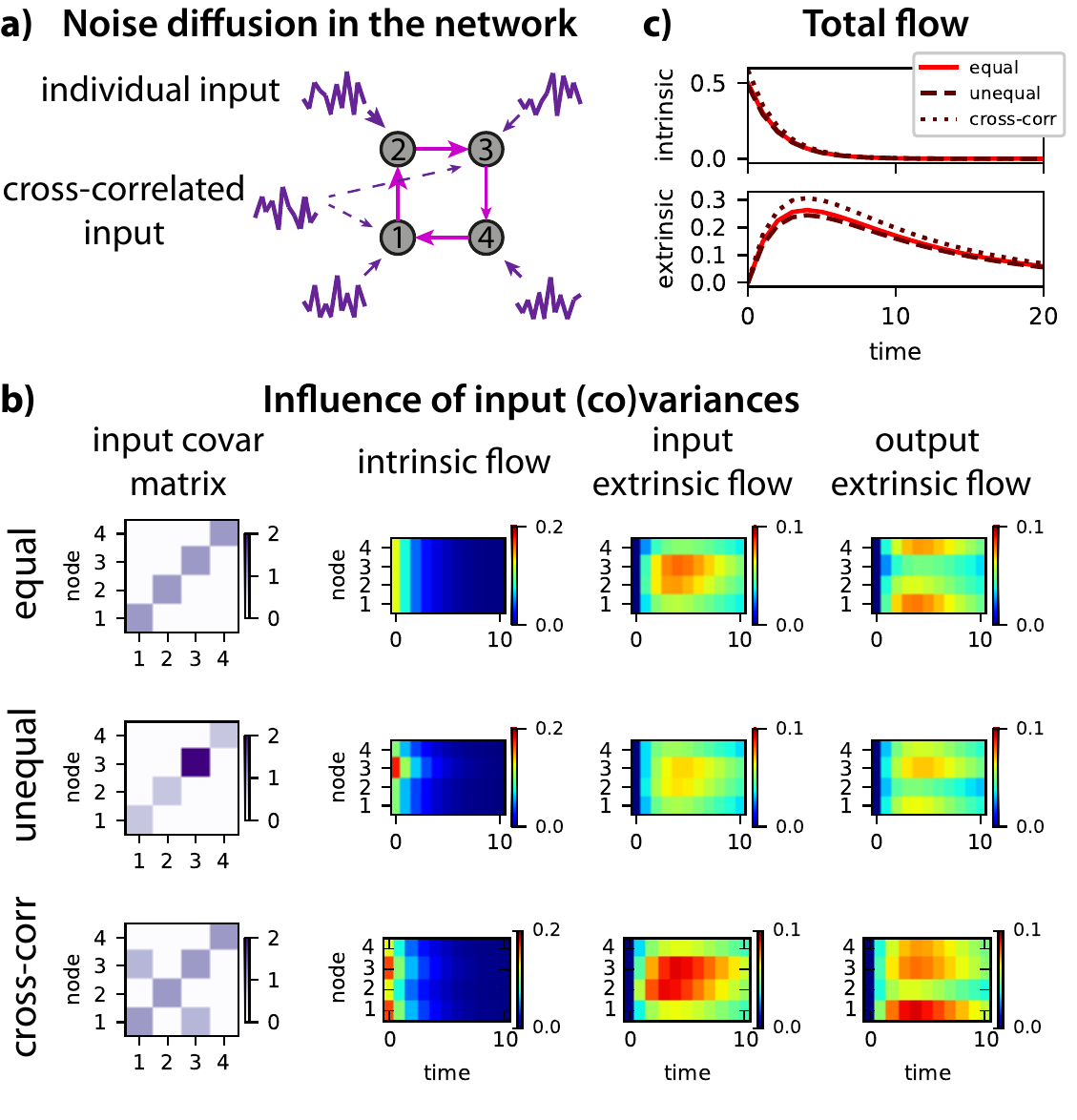}
\caption{\label{fig4_flow}
{\bf \emph{Flow} describes the noise diffusion in a MOU network.}
{\bf a)} Schematic diagram of the example network with recurrent connectivity $A$ and input covariances $\Sigma$. 
The connectivity corresponds to the magenta loop with distinct weights in \figurename~\ref{fig3_loop}.
{\bf b)} Intrinsic and (input and output) extrinsic flow for three configurations of input covariance matrices $\Sigma$ (left column).
{\bf c)} Evolution of the intrinsic and extrinsic parts of the total flow $\mathcal{S}^\mathcal{F}(t)$ over time for the three configurations in {\bf b}.
}
\end{figure*}

We have defined dynamic communicability based on the propagation kernel of the MOU process and ignoring the external noisy inputs. Now, we incorporate the input properties to describe the propagation of the fluctuating activity in the network, which fully characterizes the complex network dynamics.
This is a major theoretical novelty of our study (third dot point above): The input statistics of interest for a stable MOU process correspond to the input (co)variance matrix $\Sigma$ of the vector $\mathbf{z}$ in Eq.~\eqref{eq:MOU}, which are independent of the Jacobian $J$.
This is represented by the purple arrows with various thicknesses in \figurename~\ref{fig4_flow}a, indicating that the nodes may receive inputs with various levels of fluctuations.
When the noisy inputs received by the nodes are independent, $\Sigma$ is a diagonal matrix. 
In the general case, however, nodes may receive cross-correlated inputs (\red{spatially ``pink'' noise}), as represented by the purple dashed arrows. This corresponds to (positive) off-diagonal elements in the matrix $\Sigma$.

To quantify the propagation of this fluctuating activity, we define the flow in relation with the argument in the integral of Eq.~\eqref{eq:int_act} as
        \begin{equation} \label{eq:def_flow}
        \mathcal{F}(t) = || J^0 || e^{J t} \sqrt{\Sigma} \, ,
        \end{equation}
where $\sqrt{\Sigma}$ is the real symmetric ``square root'' matrix of the input covariance matrix, satisfying $\Sigma = \sqrt{\Sigma} \sqrt{\Sigma}^\transp$.
The spatial covariances between the node activities $Q = \langle \mathbf{x}(t) \mathbf{x}(t)^\transp \rangle$, where the angular brackets $\langle \cdot \rangle$ denote the averaging over randomness induced by $\mathbf{z}$--- can be rewritten in terms of the integrated flow over time~\cite{Lutkepohl2005} as
        \begin{equation}
        Q = \int_{0}^{+\infty} e^{J t} \, \Sigma \, e^{J^\transp t} \dd t
        = || J^0 ||^{-2} \int_{0}^{+\infty} \mathcal{F}(t) \, [\mathcal{F}(t)]^\transp \dd t \, .
        \end{equation}
In other words, our definition of flow $\mathcal{F}(t)$ can be thought as the ``square root'' of the correlation of the propagating noise, whose integration over time is the zero-lag covariance matrix $Q$.
Note that the entropy of $Q$ was used to define network complexity~\cite{Tononi1994, Galan2008, Barnett_PRE_2009}.
The flow in Eq.~\eqref{eq:def_flow} can be decomposed into two components, one related to the leakage and one to the network effect induced by the recurrent connectivity (related to communicability), so that 
        \begin{eqnarray} \label{eq:intr_extr_flow}
        \mathcal{F}^\mathrm{intr}(t) & = & || J^0 || e^{J^0 t} \sqrt{\Sigma} \ ,
        \\
        \mathcal{F}^\mathrm{extr}(t) & = & \mathcal{C}(t) \sqrt{\Sigma} \ .
        \nonumber
        \end{eqnarray}
	
Interpreting the MOU process as a noise-diffusion network~\cite{Gilson_BiolCyb_2017}, the diagonal elements of $\Sigma$ represent the amount of fluctuating intrinsic activity to the nodes, whose total determines the overall input to the network.
This means that configurations of $\sqrt{\Sigma}$ corresponding to $\Sigma$ with the same trace (sum of diagonal elements) inject the same amount of input ``noise'' in the network.
By adjusting the diagonal of $\sqrt{\Sigma}$, we can redistribute the propagation of fluctuating activity injected to the network nodes, which modulates the total flow $\mathcal{S}^\mathcal{F}(t)$ at each time, as illustrated in \figurename~\ref{fig4_flow}b and c where the intrinsic and extrinsic flows are represented separately.
\figurename~\ref{fig4_flow} shows that the intrinsic part, although initially larger, quickly becomes smaller than the extrinsic part.
Therefore, the extrinsic part is more important for the long-term behavior and the network pattern of interactions between nodes.
Note also that the same normalization is kept as with $\mathcal{C}(t)$.
Compared to the first $\Sigma$ configuration with equal excitabilities for all nodes (red solid curves in \figurename~\ref{fig4_flow}c), the second configuration sets larger excitability for node 3 with low output strength, which slightly decreases the total extrinsic flow (dark-red dashed curves).
Last, cross-correlated inputs correspond to synergetic inputs (with cross-correlations) for the MOU dynamic model and induce an extra contribution to the flow, which can strongly affect the extrinsic flow as shown with the bottom configuration in \figurename~\ref{fig4_flow}b. 
Off-diagonal elements of $\Sigma$ induce a superlinear contribution to the flow $\mathcal{F}(t)$.
This is further illustrated in \figurename~\ref{fig4_flow}c where the dark-red dotted curves are above the others, especially for the extrinsic flow.
In the following we concentrate on the \emph{extrinsic flow} $\mathcal{F}^\mathrm{extr}(t)$ in Eq.~\eqref{eq:intr_extr_flow}, which we simply refer to as \emph{flow} $\mathcal{F}(t)$.

It is worth noting from the difference in the bottom row of \figurename~\ref{fig4_flow}c that the structure of the output flow is affected by changes in $\Sigma$ for the corresponding nodes (1 and 3), whereas changes in the input flow concern the whole network.
This can be understood from Eq.~\eqref{eq:def_flow} which defines a linear mapping for $\sqrt{\Sigma}$: A change in cross-correlations for given inputs changes the corresponding columns in $\sqrt{\Sigma}$, which in turn affect the same columns in $\mathcal{F}(t)$.
To further illustrate the effect of input cross-correlations, we consider a toy example with two nodes. Preserving the diagonal of $\Sigma$ corresponds to the constraint on $a^2+b^2$ to be constant, namely comparing
        \begin{eqnarray}
        \sqrt{\Sigma}
        & = &
        \begin{pmatrix}
        a & b \\
        b & a
        \end{pmatrix}
        \quad \textrm{for} \quad 
        \Sigma
        =
        \begin{pmatrix}
        a^2+b^2 & 2 a b \\
        2 a b & a^2+b^2
        \end{pmatrix}
        \\
        \textrm{against} \quad 
        \sqrt{\Sigma'}
        & = &
        \begin{pmatrix}
        a' & 0 \\
        0 & a'
        \end{pmatrix}
        \quad \textrm{for} \quad 
        \Sigma'
        =
        \begin{pmatrix}
        a'^2 & 0 \\
        0 & a'^2
        \end{pmatrix}
        \quad \textrm{with} \quad a'^2 = a^2+b^2
        \nonumber \, .
        \end{eqnarray}
The extra contribution to the extrinsic part of $\mathcal{F}(t)$ due to the cross-correlations $b$ is thus determined by $a + b - a' = a + b - \sqrt{a^2+b^2} > 0$ when the inputs are positively correlated ($b > 0$) multiplied by the corresponding two columns of the communicability $\mathcal{C}(t)$; conversely, the contribution is negative for negative cross-correlations. 
In conclusion, synergistic inputs induce an increase of flow, which is consistent with previous definitions~\cite{Stramaglia2014}.

\subsection{Definition of diversity $\mathcal{D}$ and spectral properties of the flow}

Now we define the \emph{diversity} of the matrices $\mathcal{F}(t)$ or $\mathcal{C}(t)$, which can be seen as a proxy for the rearrangement of the node interaction structure over time.
Diversity $\mathcal{D}$ is a time-dependent measure defined as a coefficient of variation:
        \begin{equation} \label{eq:diversity_flow}
        \mathcal{D}^\mathcal{F}(t) = \frac{\sigma_{\{i,j\}}[ \mathcal{F}_{ij}(t) ]}{\mu_{\{i,j\}}[ \mathcal{F}_{ij}(t) ]} \, ,
        \end{equation}
where $\mu_{\{i,j\}}$ and $\sigma_{\{i,j\}}$ are the mean and standard deviation over the matrix elements indexed by $(i,j)$. 
The same definition holds for $\mathcal{C}(t)$.

The spectrum of the Laplacian in Eq.~\eqref{eq:Laplacian} plays a major role in exploring the hierarchical structure of networks~\cite{Arenas2006, Schaub2012}, as illustrated in \figurename~\ref{fig1_intro}c.
Assuming that the Jacobian $J$ is diagonalizable~\cite{Bernstein2009}, one can write $J = P \Lambda P^{-1}$, where $\Lambda$ is a diagonal matrix, the columns of $P$ are the right eigenvectors $\mathbf{v}^k$ of $J$, while the rows of $P^{-1}$ are the left eigenvectors $\mathbf{u}^k$, thus forming a dual basis of $\mathbf{v}^k$.
For an identical $\tau$ for all nodes, communicability can be expressed in terms of the eigenvalues $\lambda_k$ (on the diagonal of $\Lambda$) and their associated right/left eigenvectors:
        \begin{equation} \label{eq:eigen_comm}
        \mathcal{C}(t) = \sum_{1 \leq k \leq N} \frac{e^{-t/\tau} (e^{\lambda_k t} - 1)}{N \tau} \mathbf{v}^k \mathbf{u}^{k \transp} \ .
        \end{equation}
Note that a similar spectral decomposition is presented in Eq.~\eqref{eq:decomp_Laplacian} for the Laplacian, as a comparison.
The larger the real part of $\lambda_k$, the later in time is the peak for the time-dependent function $e^{-t/\tau} (e^{\lambda_k t} - 1)$ in Eq.~\eqref{eq:eigen_comm} and the larger its maximum. This means that small eigenvalues are expressed first, but weakly, while the dominating eigenvalues (with smallest negative real part) correspond to late large peaks. 
Eigenvalues with non-zero imaginary part induce damped oscillations over time.
Switching from communicability to flow in the former calculations simply implies the replacement of the left eigenvectors $\mathbf{u}^{k}$ by $\sqrt{\Sigma} \mathbf{u}^{k}$ to take into account the input statistics $\Sigma$:
        \begin{equation} \label{eq:eigen_flow}
        \mathcal{F}(t) = \sum_{1 \leq k \leq N} \frac{e^{-t/\tau} (e^{\lambda_k t} - 1)}{N \tau} \mathbf{v}^k \big(\sqrt{\Sigma} \mathbf{u}^{k} \big)^\transp \ .
        \end{equation}

Assuming that a the connectivity $A$ corresponds to a spectrum with a dominating eigenvalue $\lambda_\mathrm{max}$.
The corresponding eigenvectors are $\mathbf{v}^\mathrm{max} \simeq \mathbf{e} / \sqrt{N}$ and $\mathbf{u}^\mathrm{max} \simeq \mathbf{e} / \sqrt{N}$, which was used in Eq.~\eqref{eq:tot_comm_av} to evaluate the mean communicability. For the flow, this becomes
        \begin{eqnarray} \label{eq:mean_flow_spec}
	\mu_{\{i,j\}}[\mathcal{F}_{ij}(t)]
	& = &
	\frac{\mathcal{S}^\mathcal{F}(t)}{N^2}
	=
        \sum_{1 \leq k \leq N} \frac{e^{-t/\tau} (e^{\lambda_k t} - 1)}{N \tau} \frac{\mathbf{e}^\transp \mathbf{v}^{k} \big(\sqrt{\Sigma} \mathbf{u}^{k} \big)^\transp \mathbf{e}}{N^2}
        \\
	& = &
	\frac{e^{-t/\tau} (e^{\lambda_\mathrm{max} t} - 1)}{N \tau} 
	\Bigg[ \frac{\underline{\sqrt{\Sigma}}}{N^2} + o(1) \Bigg]
	\nonumber\ .
        \end{eqnarray}
We have defined the sum of matrix elements $\underline{\sqrt{\Sigma}} = \mathbf{e}^\transp \sqrt{\Sigma} \mathbf{e} = \sum_{1 \leq i,j \leq N} \sqrt{\Sigma}_{ij}$ and $o(1)$ indicates that the effect of all other eigenvalues than the dominating one vanishes quickly in comparison. 
However, deviations from this expected average arise from ``irregularities'' in the connectivity (e.g., due to its sparsity) and are reflected in the second-order statistics of the flow across all node pairs:
        \begin{eqnarray} \label{eq:var_flow_spec}
        \mathcal{F}(t) [\mathcal{F}(t)]^\transp
        & = &
        \sum_{1 \leq k,l \leq N} \frac{e^{-t/\tau} (e^{\lambda_k t} - 1)}{N \tau} \frac{e^{-t/\tau} (e^{\lambda_l t} - 1)}{N \tau}
        \mathbf{v}^{k} \mathbf{u}^{k \transp} \Sigma \mathbf{u}^{l} \mathbf{v}^{l \transp}
        \\
        & = &
	\Bigg[ \frac{e^{-t/\tau} (e^{\lambda_\mathrm{max} t} - 1)}{N \tau} \Bigg]^2
	\mathbf{v}^{\mathrm{max}} \mathbf{u}^{\mathrm{max} \transp} \Sigma \mathbf{u}^{\mathrm{max}} \mathbf{v}^{\mathrm{max} \transp}
	\nonumber\\
	& & +
	\frac{e^{-t/\tau} (e^{\lambda_\mathrm{max} t} - 1)}{N \tau}
	\Bigg[ \sum_{k \neq \mathrm{max}} \frac{e^{-t/\tau} (e^{\lambda_k t} - 1)}{N \tau} \mathbf{v}^{\mathrm{max}} \mathbf{u}^{\mathrm{max} \transp} \Sigma \mathbf{u}^{k} \mathbf{v}^{k \transp} + o(1) \Bigg]
	\nonumber\ .
        \end{eqnarray}
The standard deviation over the matrix elements can be evaluated using the trace of the matrix in Eq.~\eqref{eq:var_flow_spec}
        \begin{equation} \label{eq:std_flow_spec}
        \sigma_{\{i,j\}}[\mathcal{F}(t)]
        =
        \sqrt{\mathrm{tr}\Big[ \mathcal{F}(t) [\mathcal{F}(t)]^\transp \Big]}
        =
		\frac{e^{-t/\tau} (e^{\lambda_\mathrm{max} t} - 1)}{N \tau}
		\Bigg[ \frac{\sqrt{\mathrm{tr} \big( \mathbf{v}^{\mathrm{max}} \mathbf{v}^{\mathrm{max} \transp} \big)} \;
		\sqrt{\mathbf{u}^{\mathrm{max} \transp} \Sigma \mathbf{u}^{\mathrm{max}}}}{N^2} + o(1) \Bigg]
		\ .
        \end{equation}
In the end, $\mathcal{D}$ depends differently on the left and right eigenvectors:
        \begin{equation} \label{eq:heter_spec}
        \mathcal{D}^\mathcal{F}(t)
        =
	\frac{ \sqrt{\mathrm{tr} \big( \mathbf{v}^{\mathrm{max}} \mathbf{v}^{\mathrm{max} \transp} \big)} \;  \sqrt{\mathbf{u}^{\mathrm{max} \transp} \Sigma \mathbf{u}^{\mathrm{max}}}}{\underline{\sqrt{\Sigma}}} + o(1)
        \ ,
        \end{equation}
where $o(1)$ lumps together the terms in the last line of Eq.~\eqref{eq:var_flow_spec}, which decay exponentially as $e^{(\lambda_k - \lambda_\mathrm{max}) t / 2}$ for the corresponding eigenvalues (real or with imaginary parts). The same phenomenon as in \figurename~\ref{fig1_intro}c is at work here: Eigenvalues close to the dominating one(s) have a longer-lasting effect. Diversity $\mathcal{D}$ is thus predicted to converge to a non-zero asymptotic value, with a speed of convergence depending on the spectrum of $A$.

\section{Benchmark of communicability and flow using synthetic networks} \label{sec3}

The previous section has established a theoretical framework to characterize complex network dynamics. In this section we show how this extends the classic approach of graph measures, which aims to extract information about the network topology.
To do so, we base our network analysis on the (extrinsic) \emph{flow} in Eq.~\eqref{eq:intr_extr_flow}, simply denoted by $\mathcal{F}(t)$, respectively. When inputs are ignored, we sometimes employ \emph{(dynamic) communicability} $\mathcal{C}(t)$, as they coincide.
More precisely, we show how the time-dependent measures of total flow $\mathcal{S}$ and flow diversity $\mathcal{D}$ in Eq.~\eqref{eq:diversity_flow} can be used to compare networks dynamics and, beyond, compare networks. From the equations above, the intuitive interpretation is that the total flow $\mathcal{S}$ reflects the global network feedback (sum of all interactions between nodes at a given time). In contrast, $\mathcal{D}$ measures the heterogeneity of those interactions (as a coefficient of variation).
In addition, we investigate the functional roles of the nodes (e.g., feeders and receivers) that can be studied via the input/output communicability and flow, as suggested in \figurename~\ref{fig3_loop}b and \figurename~\ref{fig4_flow}b.

Practically, we examine in depth the behavior of these measures in several benchmark networks. 
We begin with randomly connected networks to understand the effect of the size, density and mean weight. 
Then we examine small-world ring lattices and hierarchical networks to uncover the interplay between the connectivity and input properties.
For these three well-known examples from graph theory, we consider directed and/or weighted networks.
Finally, we study an last example from dynamic systems, balanced excitatory-inhibitory networks.
In each case, we will illustrate the practical use of the tools introduced in Section~\ref{sec2}.

\subsection{Communicability and flow capture the properties of the network interactions and inputs}

\begin{figure*}[t!]
\includegraphics[scale=1]{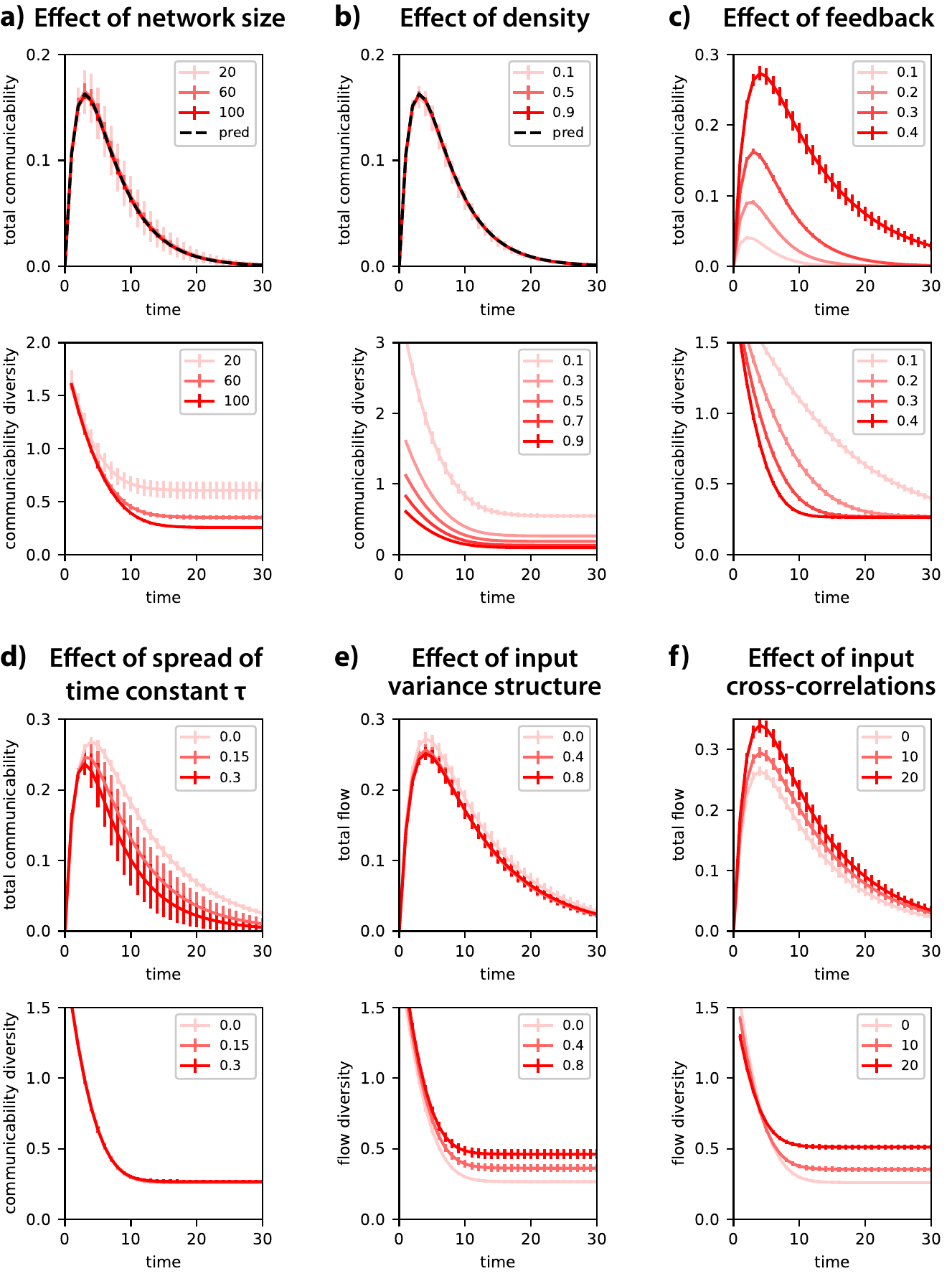}
\caption{\label{fig5_random}
{\bf Dynamic communicability and flow in randomly connected networks.}
Influence of properties of the dynamic system on the total communicability/flow $\mathcal{S}$ (top panels) and its diversity $\mathcal{D}$ (bottom panel) in a random network: {\bf a)} network size; {\bf b)} connectivity density (while preserving the mean input strength); {\bf c)} mean input weight per node; {\bf d)} spread of distribution of $\tau_i$ on the diagonal of the Jacobian in Eq.~\eqref{eq:Jacobian}; {\bf e)} colinearity between the input variances and the dominating left eigenvector of the connectivity, related to $\mathbf{u}^{\mathrm{max} \transp} \Sigma \mathbf{u}^{\mathrm{max}}$ in Eq.~\eqref{eq:heter_spec}; and {\bf f)} number of cross-correlated inputs.
The error bars correspond to the variability over 10 simulated networks. The dashed black curves in {\bf a-b} come from Eq.~\eqref{eq:tot_comm_av}. In {\bf c-f}, the networks have 100 nodes with 30\% density and same mean input weight per node.
}
\end{figure*}

As a first example, we consider randomly connected graphs. For the adjacency matrix $A$, the dominating eigenvalue $\lambda_\mathrm{max}$ is determined by the average input weight $A^\mathrm{in}$ to each node and the remaining eigenvalues are distributed around zero. 
We illustrate using numerical simulations how the network properties and the influence of the diagonal elements $-1/\tau_i$ are captured by the total flow $\mathcal{S}^\mathcal{F}(t)$ ---equal to $\mathcal{S}^\mathcal{C}(t)$ in the case of uniform inputs--- and its diversity $\mathcal{D}^\mathcal{F}(t)$.
The normalization by $|| J^0 ||^{-1}$ in Eq.~\eqref{eq:def_comm} allows for the comparison of network with various sizes in Eqs.~\eqref{eq:tot_comm_av} and \eqref{eq:mean_flow_spec}, as illustrated in the top panel of \figurename~\ref{fig5_random}a. 
This figure shows a finite size effect where the diversity $\mathcal{D}^\mathcal{C}(t)$ of smaller networks stabilizes at larger values (i.e., larger noise in $\mathbf{u}^\mathrm{max}$ compared to $\mathbf{e}/\sqrt{N}$). 
In contrast, increasing the density reduces the variability homogeneously across time.
Interestingly, \figurename~\ref{fig5_random}c shows that a weaker network feedback shortens the response, but delays the homogenization of communicability. 
The mean input weight per node is thus the main factor regulating the homogenization speed for the nodal activities in random networks, unlike the network size and density in \figurename~\ref{fig5_random}a-b. 
Using heterogeneous time constants $\tau_i$ (randomly distributed with various spreads in \figurename~\ref{fig5_random}d) induces an overall stronger leakage compared to homogeneous $\tau_i = \tau$ at the mean value, which weakens the total communicability $\mathcal{S}$.
In these four cases, the curve for $\mathcal{D}^\mathcal{C}(t)$ exhibits the predicted decay over time, which comes from all eigenvalues compared to the dominating eigenvalue. 
Note that the phenomenon is similar to \figurename~\ref{fig1_intro}c for the Laplacian with 0 as dominating eigenvalue.

Last, we vary the input properties and examine the resulting flow. 
In \figurename~\ref{fig5_random}e, we adjust the distribution of the input variances on the diagonal of $\Sigma$ to reproduce the structure of the dominating left eigenvector $\mathbf{u}^{\mathrm{max}}$ (0 means identical variances and larger coefficients indicate stronger colinearity). 
This confirms that the asymptotic diversity comes from the structure of the dominating left eigenvectors $\mathbf{u}^\mathrm{max}$ in Eq.~\eqref{eq:heter_spec}. 
A similar tuning of $\Sigma$ with respect to $\mathbf{v}^\mathrm{max}$ does not affect the diversity.
Moreover, positive input cross-correlations between nodes increase the total flow $\mathcal{S}$, as depicted for the example in \figurename~\ref{fig4_flow}c; we observe in \figurename~\ref{fig5_random}f that they also increase the asymptotic level of diversity.
From all results in \figurename~\ref{fig5_random}, 
In conclusion, connectivity properties that leave the mean input weight per node unchanged do not modify the convergence speed of the diversity $\mathcal{D}$. The latter is not affected by the input properties either.

\subsection{Interplay between local connectivity, long-range connectivity and inputs in ring lattices}

\begin{figure*}
\includegraphics[scale=0.85]{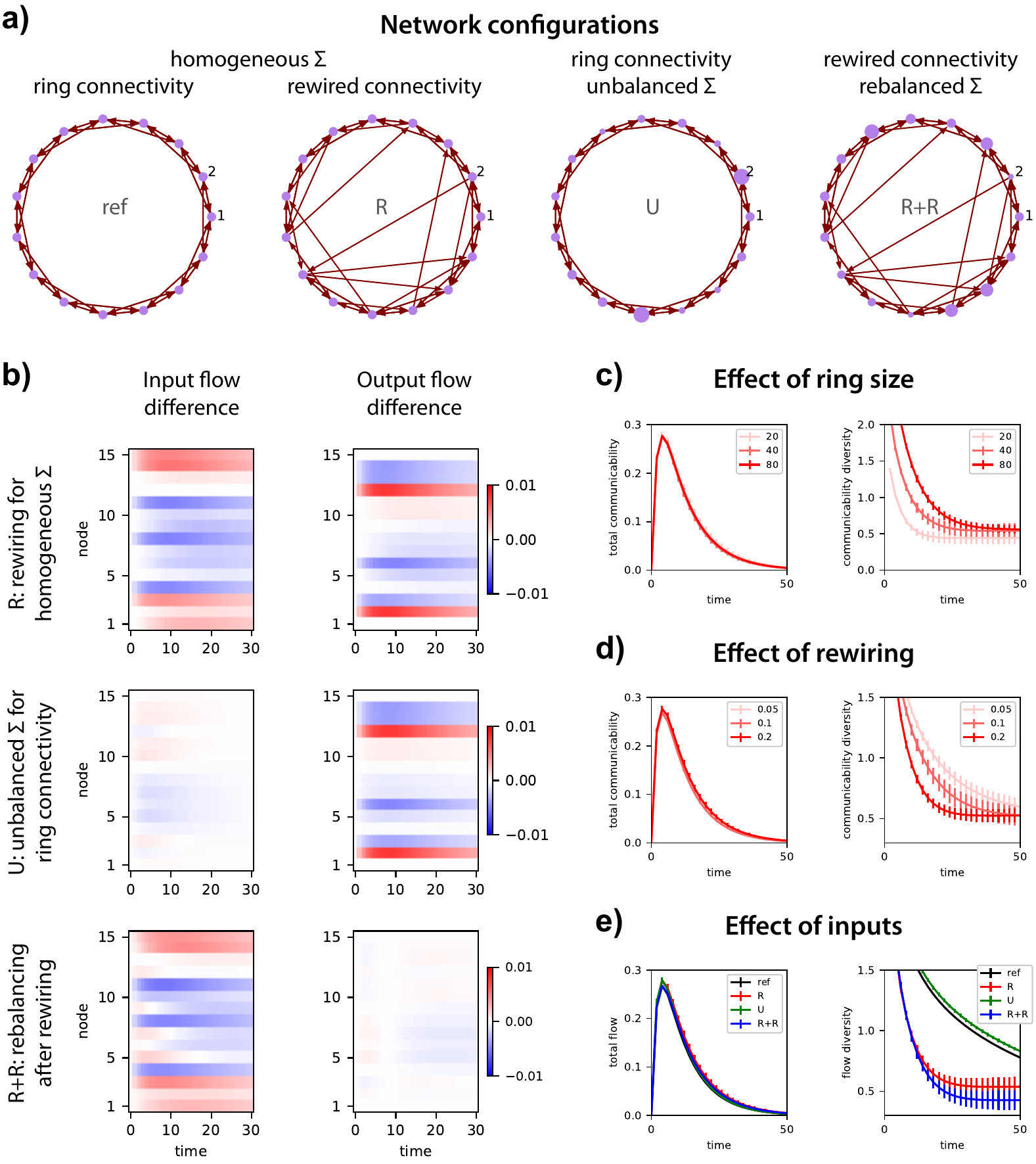}
\caption{\label{fig6_lattice}
{\bf Flow in a ring lattice.}
{\bf a)} Reference ring lattice with $N = 15$ nodes (``ref'', left diagram) and three reconfigurations. In the original ring (`ref'), each node is connected bidirectionally up to its second neighbor (4 connections per node). For configuration `R', the connectivity differ by 7 rewired connections, with directed connections from sources 2, 6, 9, 10, 12 and 13 to targets 3, 4, 8, 9, 10, 14 and 15. The two left configurations have homogeneous input variances (diagonal of $\Sigma$), as indicated by the sizes of the purple circles. Configurations `U' and `R+R' are copies of the same two connectivities, but have distinct variances. In each graph, the first two nodes are labelled for reference.
{\bf b)} Differences in input and output flow (top and bottom panels, resp.) for each of the three transformations with respect to the reference configuration in panel {\bf a}. For `U', $\Sigma$ was adjusted to mimic the effect of the rewired connectivity (`R', with homogeneous $\Sigma$) on the output flow. For `R+R', $\Sigma$ was set to obtain a similar output flow as with the ring lattice `ref', yielding a weak difference. See the main text for detail.
{\bf c)} Total communicability $\mathcal{S}$ and diversity $\mathcal{D}$ for 20 networks with 20\% rewiring of various sizes.
{\bf d)} Same as in {\bf c} with networks of size $N = 40$ and three rewiring ratios.
{\bf e)} Same as in {\bf d} for the flow with the reference and transformed networks in {\bf a-b}.
}
\end{figure*}

Now we focus on a particular network topology for which there is an implicit notion of distance between nodes, a ring lattice. For the left network in \figurename~\ref{fig6_lattice}a, all nodes have the same connectivity and input properties, so that the corresponding flow is homogeneous. 
From this original configuration, we alter the connectivity by rewiring a number of connections, resulting in long-range connections that increase the ``small-world'' property of the network.
For configuration ``R'' in \figurename~\ref{fig6_lattice}a, node 8 has an additional incoming connection, which corresponds to an expected increase of input flow in \figurename~\ref{fig6_lattice}b (top panel). 
In contrast, weakened local connectivity (with missing links between nodes 9 and 10) results in smaller input communicability for all neighbors (nodes 4 to 11), as compared to the initial ring. 
The input and output flows thus provide a proper quantification for the roles of the nodes in broadcasting and listening to the rest of the network, which combines the local and long-range connections.
Note that all nodes have identical inputs, so the flow is equal to communicability.

Rewiring does not affect the mean feedback, which leaves the total communicability $\mathcal{S}^\mathcal{C}(t)$ unchanged (left panel in \figurename~\ref{fig6_lattice}d).
Changing the network size has no effect either (left panel in \figurename~\ref{fig6_lattice}c).
However, the right panels in \figurename~\ref{fig6_lattice}c-d show the opposing effects of the ring size and rewiring probability upon the diversity $\mathcal{D}^\mathcal{C}(t)$: Larger rings take more time to homogenize (unlike random networks), but enhancing the ``small-world'' property by rewiring fastens the homogenization. 
These properties also slightly affect the asymptotic value of $\mathcal{D}$. 
In the rewired ring lattices, the input communicability is determined by the input degree (Pearson coefficient of 0.95 with p-value~$\sim 0$), unlike the output communicability (Pearson coefficient around 0 with p-value~$> 0.1$).

We also change the properties of the inputs (as indicated by the node sizes in the two right panels of \figurename~\ref{fig6_lattice}a) to investigate the combined effects on the flow. 
The rows in \figurename~\ref{fig6_lattice}b compare the deformations of the input and output flows induced by the three network modifications.
With original ring connectivity, it is possible to adjust the inputs $\Sigma$ to obtain a very similar output flow to that for rewiring, as can be seen by comparing configurations ``R'' and ``U'' in \figurename~\ref{fig6_lattice}b. 
The procedure consists in constructing a diagonal $\Sigma$ such that $\sqrt{\Sigma}$ has the desired nodal profile of output communicability evaluated at the peak of the total communicability $\mathcal{S}$ to mimic. 
Nevertheless, the input flows of the corresponding left panels differ strongly.
In the bottom row ``R+R'', we use the same trick of tuning the inputs $\Sigma$ such that the output flow of the rewired network resembles the output flow of the original homogeneous ring ``ref''.
Interestingly, nodes 5, 7 and 9 exhibit an initial increase followed by a decrease for the input flow, indicating multiple timescales.
These examples show the increased complexity of the dynamics resulting from the combined heterogeneous inputs and heterogeneous connectivity. 
Finally, Fig~\ref{fig6_lattice}e illustrates the influence of the unbalanced/rebalanced inputs upon $\mathcal{D}$ for the three reconfigurations in Fig~\ref{fig6_lattice}a performed on 20 networks: In one case they weaken the homogenization (green versus black), or conversely strengthen it (blue versus red). 
This shows that the input properties determine the asymptotic values, but only weakly affect the convergence speed.

\subsection{Community merging in hierarchical networks}

\begin{figure*}[t!]
\includegraphics[scale=1]{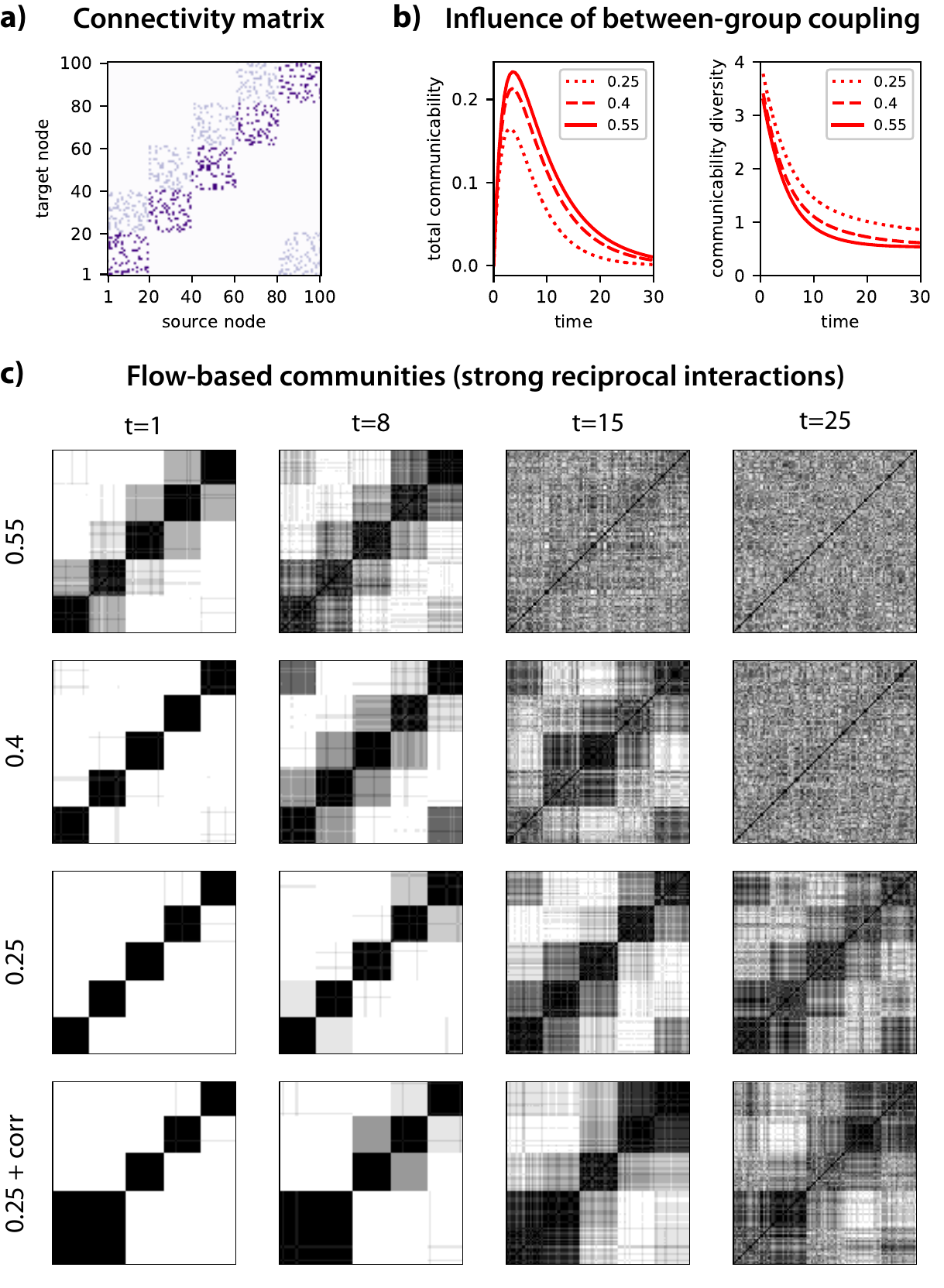}
\caption{\label{fig7_hier}
{\bf Community analysis for the loop of 5 groups.}
{\bf a)} Example of adjacency matrix with 5 groups of 20 nodes with strong random connectivity (diagonal blocks). In addition, the groups are connected as a loop (off-diagonal blocks). 
In the following we vary the common strength of the between-group connections, while keeping fixed the weight of the within-group connections.
{\bf b)} Influence of the between-group weight ---indicated in the legend as a fraction of the within-group weights--- on the total communicability $\mathcal{S}$ and its diversity $\mathcal{D}$.
{\bf c)} Communities detected employing Newman's modularity greedy algorithm on the flow $\mathcal{F}(t)$ at 4 time snapshots $t$.
The plotted values represent the (averaged over 10 simulations) participation indices for each pair of nodes. 
Black indicates that two nodes are always in the same community. 
The node ordering is the same as in {\bf a}.
The simulated networks of each row differ by the between-group weight (same ratio as in {\bf b}, as indicated on the left). 
Finally, the networks of the bottom row have positively correlated inputs between group 1 and 2.
}
\end{figure*}

Here we examine the flow in hierarchical modular networks, which are commonly used to test community detection. 
We consider a network of 5 random groups of 20 nodes each with random connections between them (diagonal blocks in \figurename~\ref{fig7_hier}a). 
These groups are connected to form a unidirectional loop (off-diagonal blocks in lighter color). 
By setting the ratio of the between- and within-group connectivity strength, we regulate the ``expression'' of the groups with respect to the global dynamics. 
\figurename~\ref{fig7_hier}b shows a faster homogenization for stronger between-group connectivity in addition to larger communicability, in line with the trend for the mean feedback in random networks (\figurename~\ref{fig5_random}c).

In the following we rely on Newman's greedy algorithm that was originally proposed to detect communities from the weight modularity in a graph~\cite{Newman2006}. 
Adapting it to the flow $\mathcal{F}(t)$ at a given time $t$ instead, we seek flow-based communities, in which nodes have strong bidirectional interactions.
Practically, we evaluate a null model of connectivity
	\begin{equation} \label{eq:null_mod_modularity}
	\mathcal{A}^\mathrm{null}
	=
	\frac{ \mathbf{a}^\mathrm{in} {\mathbf{a}^\mathrm{out}}^\transp }{ \mathcal{S}^A }
	\, .
    \end{equation}
This gives a matrix containing the deviations from the expected strengths for each connection, given the original input and output strengths for each node ($\mathbf{a}^\mathrm{in}$ and $\mathbf{a}^\mathrm{out}$, resp.), as well the total sum $\mathcal{S}^A$.
Then, we evaluate a null model for the flow $\mathcal{F}^\mathrm{null}$ using the expression in Eq.~\eqref{eq:intr_extr_flow} with $\mathcal{A}^\mathrm{null}$ instead of $A$.
Then we aggregate nodes ---starting from a partition where each node is a singleton community--- to form a partition of $K$ communities denoted by $S_k$ that maximize the quality function $\Phi$,
	\begin{equation}
	\Phi = \sum_{1 \leq k \leq K} \sum_{i,j \in S_k} \big( \mathcal{F}(t) - \mathcal{F}^\mathrm{null}(t) \big)_{ij} + \big( \mathcal{F}(t) - \mathcal{F}^\mathrm{null}(t) \big)_{ji} \, .
        \end{equation}
At each step of the greedy algorithm, two communities are fused such that $\Phi$ maximally increases. 
The frequency rate for each pair of nodes to be in the same community is displayed in \figurename~\ref{fig7_hier}c at 4 time snapshots $t$ and 4 network configurations. Results are averages over 10 numerical experiments.
We observe a similar merging of communities over time to that observed for the map equation~\cite{Schaub2012}. 
Here the between-group connection strength determines the timescale of the merging (strongest in the top row for the largest weight), as also captured by the diversity in \figurename~\ref{fig7_hier}b (right panel).

With input correlations are applied to nodes in groups 1 and 2 (bottom row in \figurename~\ref{fig7_hier}c), these groups are detected as a single community. 
Moreover, the binding clearly persists up to $t = 15$. 
This means that functional communities ---in the sense of mixing input information in the noise-diffusion network--- can be evaluated quantitatively from the flow with usual methods of community analysis~\cite{Newman2006,Schaub2012} to partition the matrix $\mathcal{F}(t)$.

\subsection{Multiple timescales and path selection in globally balanced excitatory-inhibitory network}

\begin{figure*}[t!]
\includegraphics[scale=1]{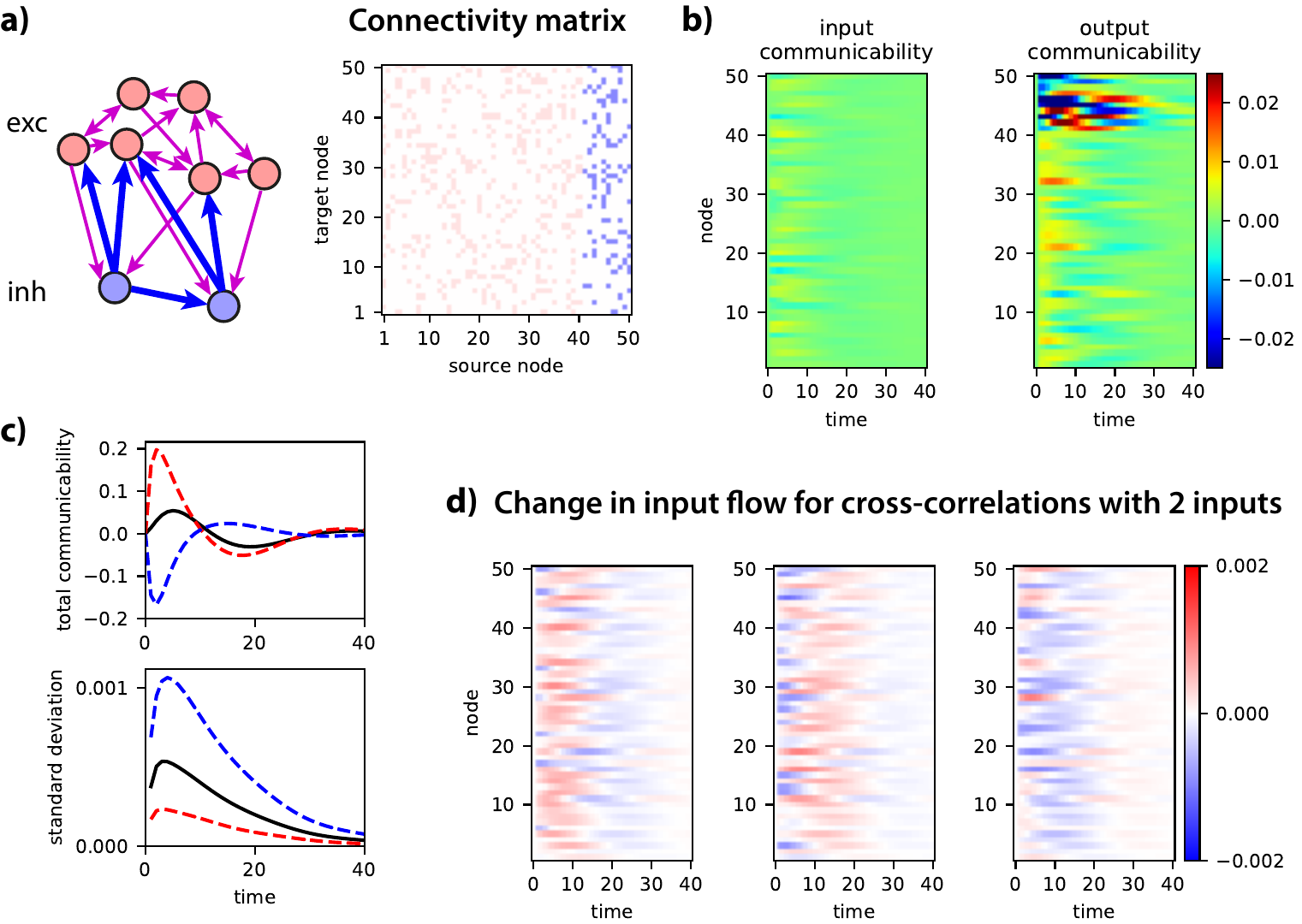}
\caption{\label{fig8_excinh}
{\bf Flow in a balanced excitatory-inhibitory network.}
{\bf a)} The network comprises 40 excitatory nodes whose outgoing connections have positive (red circles with purple arrows) and 10 inhibitory nodes with negative outgoing weights (blue circles with blue arrows). The connectivity matrix (right panel) corresponds to a density of 20\% for both types and the negative weights are 4 times larger than the positive weights to implement a global balance.
{\bf b)} Input and output communicability for the 50 nodes (41 to 50 are inhibitory).
{\bf c)} Total communicability (top panel) and its standard deviation (bottom panel) over time. The mean is indicated by the solid black curve, while the dashed red and blue curves correspond to the outgoing interactions from the excitatory and inhibitory nodes, resp.
{\bf d)} Three examples for the change in input flow due to positive cross-correlations between the same excitatory node and three distinct inhibitory nodes (one per panel).
}
\end{figure*}

In this section we discuss a case study that combines the aforementioned observations with mixed excitatory and inhibitory connections. 
We present a situation that does not usually occur in graphs, where only excitatory connections are considered; nevertheless, it can be analyzed as graphs using our framework. 
Balanced excitation and inhibition can generate extreme cases of network responses, as described by the concept of balanced amplification~\cite{Murphy2009}. 
Our purpose is to examine the heterogeneity of the communicability and flow profiles in the example network whose connectivity is represented by the matrix in \figurename~\ref{fig8_excinh}a. 
Note that the balance is global here: The positive and negative incoming weights do not exactly compensate each other for each nodes (i.e., there is no ``mass conservation'' locally), but the whole network preserves a global homogeneous steady state close to zero activity.

The balanced regime can be seen by contrasting the weak input communicability to the strong (and diverse) output communicability in \figurename~\ref{fig8_excinh}b. It is especially striking for inhibitory nodes. 
Here we have set the weights such as to obtain the dominating eigenvalue close to zero, which induces network dynamics close to the critical point where the network response diverges. 
The approximative balance between the responses from the two types of nodes is further illustrated by the total communicability $\mathcal{S}$ in \figurename~\ref{fig8_excinh}c (top panel). The bottom panel confirms the larger diversity for the inhibitory nodes. 
Note that $\mathcal{D}$ cannot be used here because the total communicability becomes zero at several points in time.

Last, we examine the flow to study the effect of correlated inputs in the balanced network. 
As said earlier, the outgoing flow is only changed for the affected nodes, but the input flow may exhibit global modifications. 
As an example, \figurename~\ref{fig8_excinh}d displays three combinations of the same excitatory node with three inhibitory nodes (one per panel): Not only are the responses of all nodes impacted, but the three situations differ vastly. 
Extrapolating, combining more than two nodes can lead to the selection of specific responses where the inputs propagate. 
These configurations may be quantitatively sought in a similar way as the output adjustments in \figurename~\ref{fig6_lattice}b for the ring network.

\section{Discussion}

Measures derived from graph theory have been increasingly used to study and compare networks estimated from real data~\cite{Borgatti2009, Sporns2000, Sporns2013, Jeong2000, Wickramasinghe2013, Kouvaris2016, Broder2000}. A usual approach is to collapse the topological information of graphs into a handful of average values (or distributions), such as the degree or clustering coefficient~\cite{Albert2002, Newman2010}, as well as make comparison with reference networks~\cite{Watts1998, Barabasi1999}.
In parallel, much effort has been dedicated to extract topological information from the emerging collective dynamics when applying specific dynamics to networks, for example based on synchrony~\cite{Arenas2008, Boccaletti2006} and random walkers~\cite{Masuda2017}. 
However, because many networks are estimated or associated with dynamic systems in real data, they should be interpreted as part of the original dynamic system, taking time into account.
To overcome this limitation, we have introduced a novel formalism based on the Green function, which allows for the analysis of the network response embodied by the interactions between nodes across time.
Although we constrain ourselves with the multivariate Ornstein-Uhlenbeck (MOU) process, corresponding to dynamics with linear feedback~\cite{Lutkepohl2005}, any network dynamics with a known or estimated Green function can be analyzed using the proposed formalism.

Our study also defines a comprehensive framework to study and compare directed and weighted networks.
A motivation of our study is to derive a canonical mathematical object that completely describes the effect of the network topology on which all subsequent analyses are based, such as interactions between nodes and community detection.
In our theory, this role is played by the \emph{dynamic communicability}. 
We have shown that many properties of the connectivity are captured by the temporal evolution of this multivariate measure (\figurename~\ref{fig5_random}a-c, \ref{fig6_lattice}c-d and \ref{fig7_hier}b).
An important aspect of our theory is that the same mathematical object is the basis of analyses at various levels (connections, communities or globally). In particular, this allows for a quantitative comparison between various network topologies.
Several time-dependent measures can also be derived to describe the roles of the nodes ---feeders or receivers--- \red{as previously done with graph communicability~\cite{Estrada2013}}. 
When considering a graph without dynamics, the corresponding dynamics for various leakage time constants ($\tau$) can be examined, which is reminiscent of the temperature in the approach proposed by~\citet{Estrada2013}.
In contrast to previous studies that use collective dynamics on networks to uncover their topological properties~\cite{Tononi1994, Arenas2006, Boccaletti2007, Rosvall2008, Galan2008, Barnett_PRE_2009, Schaub2012, Estrada2012, Estrada2013}, the link between these measures and the network dynamics is more natural here: Our analysis examines the dynamic system itself, instead of dynamics artificially applied to the network.

Another important aspect of our study is the explicit description of the propagation of external inputs to characterize the interactions between nodes, as measured by the (extrinsic) \emph{flow}.
To illustrate this point, we have revisited phenomena commonly studied in graph theory ---the small-world property in ring lattices and community merging in hierarchical modular networks--- to show the influence of inputs on them (\figurename~\ref{fig5_random}e-f, \ref{fig6_lattice}e and \ref{fig7_hier}c).
In fact, dynamic communicability is a particular case of the flow when the inputs are all identical.
In essence, the viewpoint taken on the MOU process here is that of a noise-diffusion network, where each node receives a noisy input that propagates via the network connectivity~\cite{Gilson_BiolCyb_2017}. This should be conceptually distinguished from the classical approach of the MOU process for linear regression: Here the local variabilities play the role of input variables.
The focus is thus on the second-order statistics in a stable linear-feedback system, considering the network connectivity as a transition matrix.
The concept of flow is thus important for applications in which the MOU parameters are estimated from experimental data, for both connectivity and inputs~\cite{Gilson2016, Gilson2017}.
In contrast, the equilibrium distribution for static graphs only depends on the connectivity via the Laplacian; see Eq.~\eqref{eq:stationary_distrib} in Annex~\ref{an:map_eq} for the characterization of $\tilde{\mathbf{x}}$.

In the application examples we have focused on the temporal evolution of the total flow $\mathcal{S}$ (sum of interactions) and of the corresponding diversity $\mathcal{D}$ defined as a coefficient of variation of the interactions in the network.
The total flow measures how the inputs circulate over time in the network, reflecting both the global network feedback (relatively to the leakage time constant $\tau$) and the inputs (including spatially correlated noise).
In contrast, the stabilization of the flow diversity indicates the temporal horizon when the network interactions homogenize the inputs.
It is worth noting that the flow diversity is independent of $\tau$ and can thus be used to compare the homogenization in distinct network graphs: As shown in our results, its asymptotic value reflects the heterogeneity of both the inputs and the network topology, while the speed of convergence relates to properties like small-worldness or hierarchical segregation.
Further analysis of the flow as a space-space-time tensor should be done alongside redefining and adapting classical concepts from graph theory, as was done previously when redefining graph centrality using the exponential of adjacency matrix~\cite{Estrada2007}.
Another interesting direction concerns refinements of the definition of communities, moving from non-overlapping groups with strong internal and reciprocal flow~\cite{Estrada2008} to possibly non-overlapping groups~\cite{Esquivel2011}.

\begin{acknowledgments}

MG acknowledges funding from the Marie Sk{\l}odowska-Curie Action (Grant H2020-MSCA-656547) of the European Commission.
GZL, NEK and GD acknowledge funding from the European Union's Horizon 2020 research and innovation programme under Grant Agreement No.\ 720270 (HBP SGA1). 
GD also acknowledges funding from the European Research Council Advanced Grant: DYSTRUCTURE (295129) and the Spanish Research Project (No.\ PSI2013- 42091-P).
NEK acknowledges support by the ``MOVE-IN Louvain'' fellowship co-funded by the Marie Sk{\l}odowska-Curie Action of the European Commission.

\end{acknowledgments}

\section{Annex}

\subsection{Communicability for static graph} \label{an:comm}

Communicability is a graph measure introduced by~\citet{Estrada2008} that evaluates the influence between nodes in a network and is defined as $\mathcal{C} := e^A$ for an adjacency matrix $A$. 
Using a cascade of activity in a network of propagator nodes that send a unit of activity to all connected target nodes for each unit received~\cite{Motter2002}, the network activity in continuous time obeys
	\begin{equation} 	\label{eq:cascade}
	\dot{x}_i = \sum_{1 \leq j \leq N} A_{ij} x_j \, .
	\end{equation}
Note that $A_{ij}$ is the weight from node $j$ to node $i$.
This linear cascade process obtains the solution $\mathbf{x}(t) = e^{At} \,\mathbf{x}(0)$, where $\mathbf{x}(0)$ is the initial condition. 
Therefore, $e^A$ quantifies the growth rate of the activity per unit of time $t$.
The dynamic system in Eq.~\eqref{eq:cascade} diverges for large $t$ with ``exploding'' activity because a non-trivial $A$ has at least a strictly positive eigenvalue.
The original definition~\cite{Estrada2008} has been used with directed and weighted matrices for applications with multivariate autoregressive models~\cite{Estrada2013}.

\subsection{Exponential of graph Laplacian to describe multiscale structure} \label{an:map_eq}

The map equation was first defined for discrete-time random walks in a network~\cite{Rosvall2008}. 
Later, it was formalized using the continuous-time Laplacian dynamics to describe the probability transition from node to node~\cite{Schaub2012}. 
From the adjacency matrix $A$, one can define the Laplacian matrix $L = (A-D) D^{-1}$, where $D$ is the degree matrix (a diagonal matrix containing the number of links $d_i$ of each node $i$). 
The activity $x_i$ of the node $i$ corresponds to the ratio of random walkers in this node and follows the dynamics
	\begin{equation} 	\label{eq:Laplacian}
	\dot{x}_i = -x_i + \sum_{j \neq i}\! \frac{A_{ij}}{d_j} x_j = \sum_{j=1}^N\! L_{ij} x_j \, .
	\end{equation}
The solution for this linear system is given by
	\begin{equation} 	\label{eq:exp_Lap}
	\mathbf{x}(t) = e^{L t} \mathbf{x}(0) \, .
	\end{equation}
for $t \geq 0$, which describes the (abstract) time evolution of the activity vector $\mathbf{x}$.
This system is deterministic and conservative as the sum of the presence ratios is always $\sum_i x_i = 1$. 
Note that the Laplacian $L$ is not symmetric in general, even for a symmetric $A$.
The stationary distribution $\tilde{\mathbf{x}}$ is the right eigenvector associated with the eigenvalue 0:
	\begin{equation} 	\label{eq:stationary_distrib}
	\tilde{\mathbf{x}} = L \tilde{\mathbf{x}} \, ,
	\end{equation}
which only depends on the connectivity via the Laplacian $L$.
The corresponding left eigenvector is the unit vector $\mathbf{e}^\transp$.

Now we assume that $L$ is diagonalizable and perform the same decomposition as in Eq.~\eqref{eq:eigen_comm} for the exponential matrix of the Laplacian.
In other words, $L = P \Lambda P^{-1}$ where the right eigenvectors $\mathbf{v}^{k}$ are the columns of $P$ and the left eigenvectors $\mathbf{u}^{k \transp}$ the rows of $P^{-1}$.
The Laplacian exponential can thus be written as 
	\begin{equation} 	\label{eq:decomp_Laplacian}
	e^{Lt} = P e^{\Lambda t} P^{-1} = \sum_k \! e^{\lambda_k t} \mathbf{v}^{k} \mathbf{u}^{k \transp} \, ,
	\end{equation}
where $\mathbf{v}^k$ and $\mathbf{u}^{k \transp}$ are related to the eigenvalue $\lambda_k$; the superscript $\transp$ denotes the conjugate transpose of a matrix.
This explains why $e^{Lt}$ converges toward the row matrix $\tilde{\mathbf{x}} \mathbf{e}^\transp$ as $t$ increases in \figurename~\ref{fig1_intro}c. Note that this also implies that, for any initial condition $\mathbf{x}(0)$, the activity $e^{L t} \mathbf{x}(0)$ becomes very close to the stationary distribution $\tilde{\mathbf{x}}$.


\begin{thebibliography}{45}%
	\makeatletter
	\providecommand \@ifxundefined [1]{%
		\@ifx{#1\undefined}
	}%
	\providecommand \@ifnum [1]{%
		\ifnum #1\expandafter \@firstoftwo
		\else \expandafter \@secondoftwo
		\fi
	}%
	\providecommand \@ifx [1]{%
		\ifx #1\expandafter \@firstoftwo
		\else \expandafter \@secondoftwo
		\fi
	}%
	\providecommand \natexlab [1]{#1}%
	\providecommand \enquote  [1]{``#1''}%
	\providecommand \bibnamefont  [1]{#1}%
	\providecommand \bibfnamefont [1]{#1}%
	\providecommand \citenamefont [1]{#1}%
	\providecommand \href@noop [0]{\@secondoftwo}%
	\providecommand \href [0]{\begingroup \@sanitize@url \@href}%
	\providecommand \@href[1]{\@@startlink{#1}\@@href}%
	\providecommand \@@href[1]{\endgroup#1\@@endlink}%
	\providecommand \@sanitize@url [0]{\catcode `\\12\catcode `\$12\catcode
		`\&12\catcode `\#12\catcode `\^12\catcode `\_12\catcode `\%12\relax}%
	\providecommand \@@startlink[1]{}%
	\providecommand \@@endlink[0]{}%
	\providecommand \url  [0]{\begingroup\@sanitize@url \@url }%
	\providecommand \@url [1]{\endgroup\@href {#1}{\urlprefix }}%
	\providecommand \urlprefix  [0]{URL }%
	\providecommand \Eprint [0]{\href }%
	\providecommand \doibase [0]{http://dx.doi.org/}%
	\providecommand \selectlanguage [0]{\@gobble}%
	\providecommand \bibinfo  [0]{\@secondoftwo}%
	\providecommand \bibfield  [0]{\@secondoftwo}%
	\providecommand \translation [1]{[#1]}%
	\providecommand \BibitemOpen [0]{}%
	\providecommand \bibitemStop [0]{}%
	\providecommand \bibitemNoStop [0]{.\EOS\space}%
	\providecommand \EOS [0]{\spacefactor3000\relax}%
	\providecommand \BibitemShut  [1]{\csname bibitem#1\endcsname}%
	\let\auto@bib@innerbib\@empty
	\bibitem [{\citenamefont {Borgatti}\ \emph {et~al.}(2009)\citenamefont
		{Borgatti}, \citenamefont {Mehra}, \citenamefont {Brass},\ and\ \citenamefont
		{Labianca}}]{Borgatti2009}%
	\BibitemOpen
	\bibfield  {author} {\bibinfo {author} {\bibfnamefont {S.~P.}\ \bibnamefont
			{Borgatti}}, \bibinfo {author} {\bibfnamefont {A.}~\bibnamefont {Mehra}},
		\bibinfo {author} {\bibfnamefont {D.~J.}\ \bibnamefont {Brass}}, \ and\
		\bibinfo {author} {\bibfnamefont {G.}~\bibnamefont {Labianca}},\ }\href
	{\doibase 10.1126/science.1165821} {\bibfield  {journal} {\bibinfo  {journal}
			{Science}\ }\textbf {\bibinfo {volume} {323}},\ \bibinfo {pages} {892}
		(\bibinfo {year} {2009})}\BibitemShut {NoStop}%
	\bibitem [{\citenamefont {Sporns}\ \emph {et~al.}(2000)\citenamefont {Sporns},
		\citenamefont {Tononi},\ and\ \citenamefont {Edelman}}]{Sporns2000}%
	\BibitemOpen
	\bibfield  {author} {\bibinfo {author} {\bibfnamefont {O.}~\bibnamefont
			{Sporns}}, \bibinfo {author} {\bibfnamefont {G.}~\bibnamefont {Tononi}}, \
		and\ \bibinfo {author} {\bibfnamefont {G.~M.}\ \bibnamefont {Edelman}},\
	}\href@noop {} {\bibfield  {journal} {\bibinfo  {journal} {Cereb Cortex}\
	}\textbf {\bibinfo {volume} {10}},\ \bibinfo {pages} {127} (\bibinfo {year}
	{2000})}\BibitemShut {NoStop}%
\bibitem [{\citenamefont {Sporns}(2013)}]{Sporns2013}%
\BibitemOpen
\bibfield  {author} {\bibinfo {author} {\bibfnamefont {O.}~\bibnamefont
		{Sporns}},\ }\href@noop {} {\bibfield  {journal} {\bibinfo  {journal} {Nat
			Methods}\ }\textbf {\bibinfo {volume} {10}},\ \bibinfo {pages} {491}
	(\bibinfo {year} {2013})}\BibitemShut {NoStop}%
\bibitem [{\citenamefont {Jeong}\ \emph {et~al.}(2000)\citenamefont {Jeong},
	\citenamefont {Tombor}, \citenamefont {Albert}, \citenamefont {Oltvai},\ and\
	\citenamefont {Barab{\'a}si}}]{Jeong2000}%
\BibitemOpen
\bibfield  {author} {\bibinfo {author} {\bibfnamefont {H.}~\bibnamefont
		{Jeong}}, \bibinfo {author} {\bibfnamefont {B.}~\bibnamefont {Tombor}},
	\bibinfo {author} {\bibfnamefont {R.}~\bibnamefont {Albert}}, \bibinfo
	{author} {\bibfnamefont {Z.~N.}\ \bibnamefont {Oltvai}}, \ and\ \bibinfo
	{author} {\bibfnamefont {A.~L.}\ \bibnamefont {Barab{\'a}si}},\ }\href
{\doibase 10.1038/35036627} {\bibfield  {journal} {\bibinfo  {journal}
		{Nature}\ }\textbf {\bibinfo {volume} {407}},\ \bibinfo {pages} {651}
	(\bibinfo {year} {2000})}\BibitemShut {NoStop}%
\bibitem [{\citenamefont {Wickramasinghe}\ and\ \citenamefont
	{Kiss}(2013)}]{Wickramasinghe2013}%
\BibitemOpen
\bibfield  {author} {\bibinfo {author} {\bibfnamefont {M.}~\bibnamefont
		{Wickramasinghe}}\ and\ \bibinfo {author} {\bibfnamefont {I.~Z.}\
		\bibnamefont {Kiss}},\ }\href {\doibase 10.1371/journal.pone.0080586}
{\bibfield  {journal} {\bibinfo  {journal} {PLoS One}\ }\textbf {\bibinfo
		{volume} {8}},\ \bibinfo {pages} {e80586} (\bibinfo {year}
	{2013})}\BibitemShut {NoStop}%
\bibitem [{\citenamefont {Kouvaris}\ \emph {et~al.}(2016)\citenamefont
	{Kouvaris}, \citenamefont {Sebek}, \citenamefont {Mikhailov},\ and\
	\citenamefont {Kiss}}]{Kouvaris2016}%
\BibitemOpen
\bibfield  {author} {\bibinfo {author} {\bibfnamefont {N.~E.}\ \bibnamefont
		{Kouvaris}}, \bibinfo {author} {\bibfnamefont {M.}~\bibnamefont {Sebek}},
	\bibinfo {author} {\bibfnamefont {A.~S.}\ \bibnamefont {Mikhailov}}, \ and\
	\bibinfo {author} {\bibfnamefont {I.~Z.}\ \bibnamefont {Kiss}},\ }\href
{\doibase 10.1002/anie.201607030} {\bibfield  {journal} {\bibinfo  {journal}
		{Angew Chem Int Ed Engl}\ }\textbf {\bibinfo {volume} {55}},\ \bibinfo
	{pages} {13267} (\bibinfo {year} {2016})}\BibitemShut {NoStop}%
\bibitem [{\citenamefont {Broder}\ \emph {et~al.}(2000)\citenamefont {Broder},
	\citenamefont {Kumar}, \citenamefont {Maghoul}, \citenamefont {Raghavan},
	\citenamefont {Rajagopalan}, \citenamefont {Stata}, \citenamefont {Tomkins},\
	and\ \citenamefont {Wiener}}]{Broder2000}%
\BibitemOpen
\bibfield  {author} {\bibinfo {author} {\bibfnamefont {A.}~\bibnamefont
		{Broder}}, \bibinfo {author} {\bibfnamefont {R.}~\bibnamefont {Kumar}},
	\bibinfo {author} {\bibfnamefont {F.}~\bibnamefont {Maghoul}}, \bibinfo
	{author} {\bibfnamefont {P.}~\bibnamefont {Raghavan}}, \bibinfo {author}
	{\bibfnamefont {S.}~\bibnamefont {Rajagopalan}}, \bibinfo {author}
	{\bibfnamefont {R.}~\bibnamefont {Stata}}, \bibinfo {author} {\bibfnamefont
		{A.}~\bibnamefont {Tomkins}}, \ and\ \bibinfo {author} {\bibfnamefont
		{J.}~\bibnamefont {Wiener}},\ }\href {\doibase 10.1016/S1389-1286(00)00083-9}
{\bibfield  {journal} {\bibinfo  {journal} {Comput Netw}\ }\textbf {\bibinfo
		{volume} {33}},\ \bibinfo {pages} {309} (\bibinfo {year} {2000})}\BibitemShut
{NoStop}%
\bibitem [{\citenamefont {Albert}\ and\ \citenamefont
	{Barab\'asi}(2002)}]{Albert2002}%
\BibitemOpen
\bibfield  {author} {\bibinfo {author} {\bibfnamefont {R.}~\bibnamefont
		{Albert}}\ and\ \bibinfo {author} {\bibfnamefont {A.-L.}\ \bibnamefont
		{Barab\'asi}},\ }\href {\doibase 10.1103/RevModPhys.74.47} {\bibfield
	{journal} {\bibinfo  {journal} {Rev. Mod. Phys.}\ }\textbf {\bibinfo {volume}
		{74}},\ \bibinfo {pages} {47} (\bibinfo {year} {2002})}\BibitemShut {NoStop}%
\bibitem [{\citenamefont {Newman}(2010)}]{Newman2010}%
\BibitemOpen
\bibfield  {author} {\bibinfo {author} {\bibfnamefont {M.}~\bibnamefont
		{Newman}},\ }\href@noop {} {\emph {\bibinfo {title} {Networks: An
			Introduction}}}\ (\bibinfo  {publisher} {Oxford University Press},\ \bibinfo
{year} {2010})\BibitemShut {NoStop}%
\bibitem [{\citenamefont {Bang-Jensen}\ and\ \citenamefont
	{Gutin}(2009)}]{Bang2009}%
\BibitemOpen
\bibfield  {author} {\bibinfo {author} {\bibfnamefont {J.}~\bibnamefont
		{Bang-Jensen}}\ and\ \bibinfo {author} {\bibfnamefont {G.}~\bibnamefont
		{Gutin}},\ }\href@noop {} {\emph {\bibinfo {title} {Digraphs: Theory,
			Algorithms and Applications, 2nd Edition}}}\ (\bibinfo  {publisher}
{Springer-Verlag},\ \bibinfo {year} {2009})\BibitemShut {NoStop}%
\bibitem [{\citenamefont {Barrat}\ \emph {et~al.}(2004)\citenamefont {Barrat},
	\citenamefont {Barth{\'e}lemy}, \citenamefont {Pastor-Satorras},\ and\
	\citenamefont {Vespignani}}]{Barrat2004}%
\BibitemOpen
\bibfield  {author} {\bibinfo {author} {\bibfnamefont {A.}~\bibnamefont
		{Barrat}}, \bibinfo {author} {\bibfnamefont {M.}~\bibnamefont
		{Barth{\'e}lemy}}, \bibinfo {author} {\bibfnamefont {R.}~\bibnamefont
		{Pastor-Satorras}}, \ and\ \bibinfo {author} {\bibfnamefont {A.}~\bibnamefont
		{Vespignani}},\ }\href {\doibase 10.1073/pnas.0400087101} {\bibfield
	{journal} {\bibinfo  {journal} {Proc Natl Acad Sci U S A}\ }\textbf {\bibinfo
		{volume} {101}},\ \bibinfo {pages} {3747} (\bibinfo {year}
	{2004})}\BibitemShut {NoStop}%
\bibitem [{\citenamefont {Boccaletti}\ \emph {et~al.}(2006)\citenamefont
	{Boccaletti}, \citenamefont {Latora}, \citenamefont {Moreno}, \citenamefont
	{Chavez},\ and\ \citenamefont {Hwang}}]{Boccaletti2006}%
\BibitemOpen
\bibfield  {author} {\bibinfo {author} {\bibfnamefont {S.}~\bibnamefont
		{Boccaletti}}, \bibinfo {author} {\bibfnamefont {V.}~\bibnamefont {Latora}},
	\bibinfo {author} {\bibfnamefont {Y.}~\bibnamefont {Moreno}}, \bibinfo
	{author} {\bibfnamefont {M.}~\bibnamefont {Chavez}}, \ and\ \bibinfo {author}
	{\bibfnamefont {D.-U.}\ \bibnamefont {Hwang}},\ }\href {\doibase
	https://doi.org/10.1016/j.physrep.2005.10.009} {\bibfield  {journal}
	{\bibinfo  {journal} {Physics Reports}\ }\textbf {\bibinfo {volume} {424}},\
	\bibinfo {pages} {175} (\bibinfo {year} {2006})}\BibitemShut {NoStop}%
\bibitem [{\citenamefont {Arenas}\ \emph {et~al.}(2008)\citenamefont {Arenas},
	\citenamefont {D{\'\i}az-Guilera}, \citenamefont {Kurths}, \citenamefont
	{Moreno},\ and\ \citenamefont {Zhou}}]{Arenas2008}%
\BibitemOpen
\bibfield  {author} {\bibinfo {author} {\bibfnamefont {A.}~\bibnamefont
		{Arenas}}, \bibinfo {author} {\bibfnamefont {A.}~\bibnamefont
		{D{\'\i}az-Guilera}}, \bibinfo {author} {\bibfnamefont {J.}~\bibnamefont
		{Kurths}}, \bibinfo {author} {\bibfnamefont {Y.}~\bibnamefont {Moreno}}, \
	and\ \bibinfo {author} {\bibfnamefont {C.}~\bibnamefont {Zhou}},\ }\href
{\doibase https://doi.org/10.1016/j.physrep.2008.09.002} {\bibfield
	{journal} {\bibinfo  {journal} {Physics Reports}\ }\textbf {\bibinfo {volume}
		{469}},\ \bibinfo {pages} {93} (\bibinfo {year} {2008})}\BibitemShut
{NoStop}%
\bibitem [{\citenamefont {Estrada}\ and\ \citenamefont
	{Hatano}(2008)}]{Estrada2008}%
\BibitemOpen
\bibfield  {author} {\bibinfo {author} {\bibfnamefont {E.}~\bibnamefont
		{Estrada}}\ and\ \bibinfo {author} {\bibfnamefont {N.}~\bibnamefont
		{Hatano}},\ }\href {\doibase 10.1103/PhysRevE.77.036111} {\bibfield
	{journal} {\bibinfo  {journal} {Phys Rev E Stat Nonlin Soft Matter Phys}\
	}\textbf {\bibinfo {volume} {77}},\ \bibinfo {pages} {036111} (\bibinfo
	{year} {2008})}\BibitemShut {NoStop}%
\bibitem [{\citenamefont {Bettinardi}\ \emph {et~al.}(2017)\citenamefont
	{Bettinardi}, \citenamefont {Deco}, \citenamefont {Karlaftis}, \citenamefont
	{Van~Hartevelt}, \citenamefont {Fernandes}, \citenamefont {Kourtzi},
	\citenamefont {Kringelbach},\ and\ \citenamefont
	{Zamora-L{\'o}pez}}]{Bettinardi2017}%
\BibitemOpen
\bibfield  {author} {\bibinfo {author} {\bibfnamefont {R.~G.}\ \bibnamefont
		{Bettinardi}}, \bibinfo {author} {\bibfnamefont {G.}~\bibnamefont {Deco}},
	\bibinfo {author} {\bibfnamefont {V.~M.}\ \bibnamefont {Karlaftis}}, \bibinfo
	{author} {\bibfnamefont {T.~J.}\ \bibnamefont {Van~Hartevelt}}, \bibinfo
	{author} {\bibfnamefont {H.~M.}\ \bibnamefont {Fernandes}}, \bibinfo {author}
	{\bibfnamefont {Z.}~\bibnamefont {Kourtzi}}, \bibinfo {author} {\bibfnamefont
		{M.~L.}\ \bibnamefont {Kringelbach}}, \ and\ \bibinfo {author} {\bibfnamefont
		{G.}~\bibnamefont {Zamora-L{\'o}pez}},\ }\href {\doibase 10.1063/1.4980099}
{\bibfield  {journal} {\bibinfo  {journal} {Chaos}\ }\textbf {\bibinfo
		{volume} {27}},\ \bibinfo {pages} {047409} (\bibinfo {year}
	{2017})}\BibitemShut {NoStop}%
\bibitem [{\citenamefont {Arenas}\ \emph {et~al.}(2006)\citenamefont {Arenas},
	\citenamefont {D{\'\i}az-Guilera},\ and\ \citenamefont
	{P{\'e}rez-Vicente}}]{Arenas2006}%
\BibitemOpen
\bibfield  {author} {\bibinfo {author} {\bibfnamefont {A.}~\bibnamefont
		{Arenas}}, \bibinfo {author} {\bibfnamefont {A.}~\bibnamefont
		{D{\'\i}az-Guilera}}, \ and\ \bibinfo {author} {\bibfnamefont {C.~J.}\
		\bibnamefont {P{\'e}rez-Vicente}},\ }\href {\doibase
	10.1103/PhysRevLett.96.114102} {\bibfield  {journal} {\bibinfo  {journal}
		{Phys Rev Lett}\ }\textbf {\bibinfo {volume} {96}},\ \bibinfo {pages}
	{114102} (\bibinfo {year} {2006})}\BibitemShut {NoStop}%
\bibitem [{\citenamefont {Boccaletti}\ \emph {et~al.}(2007)\citenamefont
	{Boccaletti}, \citenamefont {Ivanchenko}, \citenamefont {Latora},
	\citenamefont {Pluchino},\ and\ \citenamefont {Rapisarda}}]{Boccaletti2007}%
\BibitemOpen
\bibfield  {author} {\bibinfo {author} {\bibfnamefont {S.}~\bibnamefont
		{Boccaletti}}, \bibinfo {author} {\bibfnamefont {M.}~\bibnamefont
		{Ivanchenko}}, \bibinfo {author} {\bibfnamefont {V.}~\bibnamefont {Latora}},
	\bibinfo {author} {\bibfnamefont {A.}~\bibnamefont {Pluchino}}, \ and\
	\bibinfo {author} {\bibfnamefont {A.}~\bibnamefont {Rapisarda}},\ }\href
{\doibase 10.1103/PhysRevE.75.045102} {\bibfield  {journal} {\bibinfo
		{journal} {Phys Rev E Stat Nonlin Soft Matter Phys}\ }\textbf {\bibinfo
		{volume} {75}},\ \bibinfo {pages} {045102} (\bibinfo {year}
	{2007})}\BibitemShut {NoStop}%
\bibitem [{\citenamefont {Rosvall}\ and\ \citenamefont
	{Bergstrom}(2008)}]{Rosvall2008}%
\BibitemOpen
\bibfield  {author} {\bibinfo {author} {\bibfnamefont {M.}~\bibnamefont
		{Rosvall}}\ and\ \bibinfo {author} {\bibfnamefont {C.~T.}\ \bibnamefont
		{Bergstrom}},\ }\href {\doibase 10.1073/pnas.0706851105} {\bibfield
	{journal} {\bibinfo  {journal} {Proc Natl Acad Sci U S A}\ }\textbf {\bibinfo
		{volume} {105}},\ \bibinfo {pages} {1118} (\bibinfo {year}
	{2008})}\BibitemShut {NoStop}%
\bibitem [{\citenamefont {Tononi}\ \emph {et~al.}(1994)\citenamefont {Tononi},
	\citenamefont {Sporns},\ and\ \citenamefont {Edelman}}]{Tononi1994}%
\BibitemOpen
\bibfield  {author} {\bibinfo {author} {\bibfnamefont {G.}~\bibnamefont
		{Tononi}}, \bibinfo {author} {\bibfnamefont {O.}~\bibnamefont {Sporns}}, \
	and\ \bibinfo {author} {\bibfnamefont {G.~M.}\ \bibnamefont {Edelman}},\
}\href@noop {} {\bibfield  {journal} {\bibinfo  {journal} {Proc Natl Acad Sci
		U S A}\ }\textbf {\bibinfo {volume} {91}},\ \bibinfo {pages} {5033} (\bibinfo
{year} {1994})}\BibitemShut {NoStop}%
\bibitem [{\citenamefont {Gal{\'a}n}(2008)}]{Galan2008}%
\BibitemOpen
\bibfield  {author} {\bibinfo {author} {\bibfnamefont {R.~F.}\ \bibnamefont
		{Gal{\'a}n}},\ }\href {\doibase 10.1371/journal.pone.0002148} {\bibfield
	{journal} {\bibinfo  {journal} {PLoS One}\ }\textbf {\bibinfo {volume} {3}},\
	\bibinfo {pages} {e2148} (\bibinfo {year} {2008})}\BibitemShut {NoStop}%
\bibitem [{\citenamefont {Barnett}\ \emph {et~al.}(2009)\citenamefont
	{Barnett}, \citenamefont {Buckley},\ and\ \citenamefont
	{Bullock}}]{Barnett_PRE_2009}%
\BibitemOpen
\bibfield  {author} {\bibinfo {author} {\bibfnamefont {L.}~\bibnamefont
		{Barnett}}, \bibinfo {author} {\bibfnamefont {C.~L.}\ \bibnamefont
		{Buckley}}, \ and\ \bibinfo {author} {\bibfnamefont {S.}~\bibnamefont
		{Bullock}},\ }\href {\doibase 10.1103/PhysRevE.79.051914} {\bibfield
	{journal} {\bibinfo  {journal} {Phys Rev E Stat Nonlin Soft Matter Phys}\
	}\textbf {\bibinfo {volume} {79}},\ \bibinfo {pages} {051914} (\bibinfo
	{year} {2009})}\BibitemShut {NoStop}%
\bibitem [{\citenamefont {L\"utkepohl}(2005)}]{Lutkepohl2005}%
\BibitemOpen
\bibfield  {author} {\bibinfo {author} {\bibfnamefont {H.}~\bibnamefont
		{L\"utkepohl}},\ }\href@noop {} {\emph {\bibinfo {title} {New introduction to
			multiple time series analysis}}}\ (\bibinfo  {publisher} {Springer Science \&
	Business Media},\ \bibinfo {year} {2005})\BibitemShut {NoStop}%
\bibitem [{\citenamefont {Gilson}(2017)}]{Gilson_BiolCyb_2017}%
\BibitemOpen
\bibfield  {author} {\bibinfo {author} {\bibfnamefont {M.}~\bibnamefont
		{Gilson}},\ }\href {\doibase 10.1007/s00422-017-0741-y} {\bibfield  {journal}
	{\bibinfo  {journal} {Biological Cybernetics}\ } (\bibinfo {year} {2017}),\
	10.1007/s00422-017-0741-y}\BibitemShut {NoStop}%
\bibitem [{\citenamefont {Doob}(1942)}]{Doob1942}%
\BibitemOpen
\bibfield  {author} {\bibinfo {author} {\bibfnamefont {J.~L.}\ \bibnamefont
		{Doob}},\ }\href@noop {} {\bibfield  {journal} {\bibinfo  {journal} {Annals
			of Mathematics}\ }\textbf {\bibinfo {volume} {43}},\ \bibinfo {pages} {351}
	(\bibinfo {year} {1942})}\BibitemShut {NoStop}%
\bibitem [{\citenamefont {Vasicek}(1977)}]{Vasicek1977}%
\BibitemOpen
\bibfield  {author} {\bibinfo {author} {\bibfnamefont {O.}~\bibnamefont
		{Vasicek}},\ }\href {\doibase doi.org/10.1016/0304-405X(77)90016-2}
{\bibfield  {journal} {\bibinfo  {journal} {Journal of Financial Economics}\
	}\textbf {\bibinfo {volume} {5}},\ \bibinfo {pages} {177} (\bibinfo {year}
	{1977})}\BibitemShut {NoStop}%
\bibitem [{\citenamefont {Britton}\ and\ \citenamefont
	{Neal}(2010)}]{Britton2010}%
\BibitemOpen
\bibfield  {author} {\bibinfo {author} {\bibfnamefont {T.}~\bibnamefont
		{Britton}}\ and\ \bibinfo {author} {\bibfnamefont {P.}~\bibnamefont {Neal}},\
}\href {\doibase 10.1007/s00285-009-0320-5} {\bibfield  {journal} {\bibinfo
	{journal} {J Math Biol}\ }\textbf {\bibinfo {volume} {61}},\ \bibinfo {pages}
{763} (\bibinfo {year} {2010})}\BibitemShut {NoStop}%
\bibitem [{\citenamefont {Andersson}\ and\ \citenamefont
	{Lindenstrand}(2011)}]{Andersson2011}%
\BibitemOpen
\bibfield  {author} {\bibinfo {author} {\bibfnamefont {P.}~\bibnamefont
		{Andersson}}\ and\ \bibinfo {author} {\bibfnamefont {D.}~\bibnamefont
		{Lindenstrand}},\ }\href {\doibase 10.1007/s00285-010-0336-x} {\bibfield
	{journal} {\bibinfo  {journal} {J Math Biol}\ }\textbf {\bibinfo {volume}
		{62}},\ \bibinfo {pages} {333} (\bibinfo {year} {2011})}\BibitemShut
{NoStop}%
\bibitem [{\citenamefont {Gilson}\ \emph {et~al.}(2016)\citenamefont {Gilson},
	\citenamefont {Moreno-Bote}, \citenamefont {Ponce-Alvarez}, \citenamefont
	{Ritter},\ and\ \citenamefont {Deco}}]{Gilson2016}%
\BibitemOpen
\bibfield  {author} {\bibinfo {author} {\bibfnamefont {M.}~\bibnamefont
		{Gilson}}, \bibinfo {author} {\bibfnamefont {R.}~\bibnamefont {Moreno-Bote}},
	\bibinfo {author} {\bibfnamefont {A.}~\bibnamefont {Ponce-Alvarez}}, \bibinfo
	{author} {\bibfnamefont {P.}~\bibnamefont {Ritter}}, \ and\ \bibinfo {author}
	{\bibfnamefont {G.}~\bibnamefont {Deco}},\ }\href@noop {} {\bibfield
	{journal} {\bibinfo  {journal} {PLoS Comput Biol}\ }\textbf {\bibinfo
		{volume} {12}},\ \bibinfo {pages} {e1004762} (\bibinfo {year}
	{2016})}\BibitemShut {NoStop}%
\bibitem [{\citenamefont {Gilson}\ \emph {et~al.}(2017)\citenamefont {Gilson},
	\citenamefont {Deco}, \citenamefont {Friston}, \citenamefont {Hagmann},
	\citenamefont {Mantini}, \citenamefont {Betti}, \citenamefont {Romani},\ and\
	\citenamefont {Corbetta}}]{Gilson2017}%
\BibitemOpen
\bibfield  {author} {\bibinfo {author} {\bibfnamefont {M.}~\bibnamefont
		{Gilson}}, \bibinfo {author} {\bibfnamefont {G.}~\bibnamefont {Deco}},
	\bibinfo {author} {\bibfnamefont {K.}~\bibnamefont {Friston}}, \bibinfo
	{author} {\bibfnamefont {P.}~\bibnamefont {Hagmann}}, \bibinfo {author}
	{\bibfnamefont {D.}~\bibnamefont {Mantini}}, \bibinfo {author} {\bibfnamefont
		{V.}~\bibnamefont {Betti}}, \bibinfo {author} {\bibfnamefont {G.~L.}\
		\bibnamefont {Romani}}, \ and\ \bibinfo {author} {\bibfnamefont
		{M.}~\bibnamefont {Corbetta}},\ }\href {\doibase
	10.1016/j.neuroimage.2017.09.061} {\bibfield  {journal} {\bibinfo  {journal}
		{Neuroimage}\ } (\bibinfo {year} {2017}),\
	10.1016/j.neuroimage.2017.09.061}\BibitemShut {NoStop}%
\bibitem [{\citenamefont {Estrada}\ and\ \citenamefont
	{Hatano}(2007)}]{Estrada2007}%
\BibitemOpen
\bibfield  {author} {\bibinfo {author} {\bibfnamefont {E.}~\bibnamefont
		{Estrada}}\ and\ \bibinfo {author} {\bibfnamefont {N.}~\bibnamefont
		{Hatano}},\ }\href {\doibase 10.1016/j.cplett.2007.03.098} {\bibfield
	{journal} {\bibinfo  {journal} {Chemical Physics Letters}\ }\textbf {\bibinfo
		{volume} {439}},\ \bibinfo {pages} {247} (\bibinfo {year}
	{2007})}\BibitemShut {NoStop}%
\bibitem [{\citenamefont {Estrada}\ \emph {et~al.}(2012)\citenamefont
	{Estrada}, \citenamefont {Hatano},\ and\ \citenamefont
	{Benzi}}]{Estrada2012}%
\BibitemOpen
\bibfield  {author} {\bibinfo {author} {\bibfnamefont {E.}~\bibnamefont
		{Estrada}}, \bibinfo {author} {\bibfnamefont {N.}~\bibnamefont {Hatano}}, \
	and\ \bibinfo {author} {\bibfnamefont {M.}~\bibnamefont {Benzi}},\ }\href
{\doibase 10.1016/j.physrep.2012.01.006} {\bibfield  {journal} {\bibinfo
		{journal} {Physics Reports}\ }\textbf {\bibinfo {volume} {514}},\ \bibinfo
	{pages} {89} (\bibinfo {year} {2012})}\BibitemShut {NoStop}%
\bibitem [{\citenamefont {Schaub}\ \emph {et~al.}(2012)\citenamefont {Schaub},
	\citenamefont {Lambiotte},\ and\ \citenamefont {Barahona}}]{Schaub2012}%
\BibitemOpen
\bibfield  {author} {\bibinfo {author} {\bibfnamefont {M.~T.}\ \bibnamefont
		{Schaub}}, \bibinfo {author} {\bibfnamefont {R.}~\bibnamefont {Lambiotte}}, \
	and\ \bibinfo {author} {\bibfnamefont {M.}~\bibnamefont {Barahona}},\ }\href
{\doibase 10.1103/PhysRevE.86.026112} {\bibfield  {journal} {\bibinfo
		{journal} {Phys Rev E Stat Nonlin Soft Matter Phys}\ }\textbf {\bibinfo
		{volume} {86}},\ \bibinfo {pages} {026112} (\bibinfo {year}
	{2012})}\BibitemShut {NoStop}%
\bibitem [{\citenamefont {Mugnolo}(2017)}]{Mugnolo2017}%
\BibitemOpen
\bibfield  {author} {\bibinfo {author} {\bibfnamefont {D.}~\bibnamefont
		{Mugnolo}},\ }\href@noop {} {\bibfield  {journal} {\bibinfo  {journal}
		{arXiv}\ ,\ \bibinfo {pages} {1702.05253}} (\bibinfo {year}
	{2017})}\BibitemShut {NoStop}%
\bibitem [{\citenamefont {Estrada}(2013)}]{Estrada2013}%
\BibitemOpen
\bibfield  {author} {\bibinfo {author} {\bibfnamefont {E.}~\bibnamefont
		{Estrada}},\ }\href {\doibase 10.1103/PhysRevE.88.042811} {\bibfield
	{journal} {\bibinfo  {journal} {Phys Rev E Stat Nonlin Soft Matter Phys}\
	}\textbf {\bibinfo {volume} {88}},\ \bibinfo {pages} {042811} (\bibinfo
	{year} {2013})}\BibitemShut {NoStop}%
\bibitem [{\citenamefont {Chung}(1992)}]{Chung1992}%
\BibitemOpen
\bibfield  {author} {\bibinfo {author} {\bibfnamefont {F.~R.~K.}\
		\bibnamefont {Chung}},\ }\href@noop {} {\emph {\bibinfo {title} {Spectral
			Graph Theory}}},\ \bibinfo {number} {978-0-8218-0315-8}\ (\bibinfo
{publisher} {AMS},\ \bibinfo {year} {1992})\BibitemShut {NoStop}%
\bibitem [{\citenamefont {B{\'e}rard}\ \emph {et~al.}(1994)\citenamefont
	{B{\'e}rard}, \citenamefont {Besson},\ and\ \citenamefont
	{Gallot}}]{Berard1994}%
\BibitemOpen
\bibfield  {author} {\bibinfo {author} {\bibfnamefont {P.}~\bibnamefont
		{B{\'e}rard}}, \bibinfo {author} {\bibfnamefont {G.}~\bibnamefont {Besson}},
	\ and\ \bibinfo {author} {\bibfnamefont {S.}~\bibnamefont {Gallot}},\ }\href
{\doibase 10.1007/BF01896401} {\bibfield  {journal} {\bibinfo  {journal}
		{Geometric {\&} Functional Analysis}\ }\textbf {\bibinfo {volume} {4}},\
	\bibinfo {pages} {373} (\bibinfo {year} {1994})}\BibitemShut {NoStop}%
\bibitem [{\citenamefont {Watts}\ and\ \citenamefont
	{Strogatz}(1998)}]{Watts1998}%
\BibitemOpen
\bibfield  {author} {\bibinfo {author} {\bibfnamefont {D.~J.}\ \bibnamefont
		{Watts}}\ and\ \bibinfo {author} {\bibfnamefont {S.~H.}\ \bibnamefont
		{Strogatz}},\ }\href {\doibase 10.1038/30918} {\bibfield  {journal} {\bibinfo
		{journal} {Nature}\ }\textbf {\bibinfo {volume} {393}},\ \bibinfo {pages}
	{440} (\bibinfo {year} {1998})}\BibitemShut {NoStop}%
\bibitem [{\citenamefont {Motter}\ and\ \citenamefont
	{Lai}(2002)}]{Motter2002}%
\BibitemOpen
\bibfield  {author} {\bibinfo {author} {\bibfnamefont {A.~E.}\ \bibnamefont
		{Motter}}\ and\ \bibinfo {author} {\bibfnamefont {Y.-C.}\ \bibnamefont
		{Lai}},\ }\href {\doibase 10.1103/PhysRevE.66.065102} {\bibfield  {journal}
	{\bibinfo  {journal} {Phys Rev E Stat Nonlin Soft Matter Phys}\ }\textbf
	{\bibinfo {volume} {66}},\ \bibinfo {pages} {065102} (\bibinfo {year}
	{2002})}\BibitemShut {NoStop}%
\bibitem [{\citenamefont {Stramaglia}\ \emph {et~al.}(2014)\citenamefont
	{Stramaglia}, \citenamefont {Cortes},\ and\ \citenamefont
	{Marinazzo}}]{Stramaglia2014}%
\BibitemOpen
\bibfield  {author} {\bibinfo {author} {\bibfnamefont {S.}~\bibnamefont
		{Stramaglia}}, \bibinfo {author} {\bibfnamefont {J.}~\bibnamefont {Cortes}},
	\ and\ \bibinfo {author} {\bibfnamefont {D.}~\bibnamefont {Marinazzo}},\
}\href@noop {} {\bibfield  {journal} {\bibinfo  {journal} {New Journal of
		Physics}\ }\textbf {\bibinfo {volume} {16}},\ \bibinfo {pages} {105003}
(\bibinfo {year} {2014})}\BibitemShut {NoStop}%
\bibitem [{\citenamefont {Bernstein}(2009)}]{Bernstein2009}%
\BibitemOpen
\bibfield  {author} {\bibinfo {author} {\bibfnamefont {D.}~\bibnamefont
		{Bernstein}},\ }\href@noop {} {\emph {\bibinfo {title} {Matrix mathematics:
			Theory, facts, and formulas (second edition)}}}\ (\bibinfo  {publisher}
{PUP},\ \bibinfo {year} {2009})\BibitemShut {NoStop}%
\bibitem [{\citenamefont {Newman}(2006)}]{Newman2006}%
\BibitemOpen
\bibfield  {author} {\bibinfo {author} {\bibfnamefont {M.~E.~J.}\
		\bibnamefont {Newman}},\ }\href {\doibase 10.1073/pnas.0601602103} {\bibfield
	{journal} {\bibinfo  {journal} {Proc Natl Acad Sci U S A}\ }\textbf
	{\bibinfo {volume} {103}},\ \bibinfo {pages} {8577} (\bibinfo {year}
	{2006})}\BibitemShut {NoStop}%
\bibitem [{\citenamefont {Murphy}\ and\ \citenamefont
	{Miller}(2009)}]{Murphy2009}%
\BibitemOpen
\bibfield  {author} {\bibinfo {author} {\bibfnamefont {B.~K.}\ \bibnamefont
		{Murphy}}\ and\ \bibinfo {author} {\bibfnamefont {K.~D.}\ \bibnamefont
		{Miller}},\ }\href {\doibase 10.1016/j.neuron.2009.02.005} {\bibfield
	{journal} {\bibinfo  {journal} {Neuron}\ }\textbf {\bibinfo {volume} {61}},\
	\bibinfo {pages} {635} (\bibinfo {year} {2009})}\BibitemShut {NoStop}%
\bibitem [{\citenamefont {Barabasi}\ and\ \citenamefont
	{Albert}(1999)}]{Barabasi1999}%
\BibitemOpen
\bibfield  {author} {\bibinfo {author} {\bibfnamefont {A.}~\bibnamefont
		{Barabasi}}\ and\ \bibinfo {author} {\bibfnamefont {R.}~\bibnamefont
		{Albert}},\ }\href@noop {} {\bibfield  {journal} {\bibinfo  {journal}
		{Science}\ }\textbf {\bibinfo {volume} {286}},\ \bibinfo {pages} {509}
	(\bibinfo {year} {1999})}\BibitemShut {NoStop}%
\bibitem [{\citenamefont {Masuda}\ \emph {et~al.}(2017)\citenamefont {Masuda},
	\citenamefont {Porter},\ and\ \citenamefont {Lambiotte}}]{Masuda2017}%
\BibitemOpen
\bibfield  {author} {\bibinfo {author} {\bibfnamefont {N.}~\bibnamefont
		{Masuda}}, \bibinfo {author} {\bibfnamefont {M.~A.}\ \bibnamefont {Porter}},
	\ and\ \bibinfo {author} {\bibfnamefont {R.}~\bibnamefont {Lambiotte}},\
}\href@noop {} {\bibfield  {journal} {\bibinfo  {journal} {Phys. Reps.}\
}\textbf {\bibinfo {volume} {716--717}},\ \bibinfo {pages} {1} (\bibinfo
{year} {2017})}\BibitemShut {NoStop}%
\bibitem [{\citenamefont {Esquivel}\ and\ \citenamefont
	{Rosvall}(2011)}]{Esquivel2011}%
\BibitemOpen
\bibfield  {author} {\bibinfo {author} {\bibfnamefont {A.~V.}\ \bibnamefont
		{Esquivel}}\ and\ \bibinfo {author} {\bibfnamefont {M.}~\bibnamefont
		{Rosvall}},\ }\href@noop {} {\bibfield  {journal} {\bibinfo  {journal} {Phys
			Rev X}\ }\textbf {\bibinfo {volume} {1}},\ \bibinfo {pages} {021025}
	(\bibinfo {year} {2011})}\BibitemShut {NoStop}%
\end{thebibliography}

%

\end{document}